# Modelling the Effects of using Gas Diffusion Layers with Patterned Wettability for Advanced Water Management in Proton Exchange Membrane Fuel Cells

Jaka Dujc[1], Antoni Forner-Cuenca[2], Philip Marmet[1], Magali Cochet[2], Roman Vetter[1], Jürgen O. Schumacher*[1], and Pierre Boillat[2]

[1]*Institute of Computational Physics (ICP), Zurich University of Applied Sciences (ZHAW), Wildbachstrasse 21, CH-8401 Winterthur, Switzerland*

[2]*Electrochemistry Laboratory (LEC), Neutron Imaging and Activation Group (NIAG), Paul Scherrer Institute, 5232 Villigen PSI, Switzerland*

We present a macrohomogeneous two-phase model of a proton exchange membrane fuel cell (PEMFC). The model takes into account the mechanical compression of the gas diffusion layer (GDL), the two-phase flow of water, the transport of the gas species and the electrochemical reaction of the reactant gases. The model was used to simulate the behavior of a PEMFC with a patterned GDL. The results of the reduced model, which considers only the mechanical compression and the two-phase flow, are compared to the experimental ex-situ imbibition data obtained by neutron radiography imaging. The results are in good agreement. Additionally, by using all model features, a simulation of an operating fuel cell has been performed to study the intricate couplings in an operating fuel cell and to examine the patterned GDL effects. The model confirms that the patterned GDL design liberates the pre-defined domains from liquid water and thus locally increases the oxygen diffusivity.

## 1 Introduction

In PEMFCs water is produced by the cathode oxygen reduction reaction. The presence of liquid water in PEMFC has both positive and negative effects. On one hand water is beneficial since high water content in the membrane increases proton conductivity and thus overall fuel cell efficiency, while evaporation of water cools down the fuel cell. On the other hand the liquid water accumulates in the porous gas diffusion layer (GDL) and thus limits the transport of oxygen.

Transport properties of dry GDLs at different mechanical compression states were recently extracted from the microstructure obtained by X-ray tomographic microscopy [1]. These properties include the gas diffusivity, permeability and electrical conductivity. Transport properties at different levels of liquid water saturation were investigated by the same group in a different study [2].

Several approaches have been proposed to improve the water management by lateral patterning of GDLs: the common feature of all these approaches is a modification of the GDLs in order to pre-define the water removal pathways. In the "perforation approach" (e.g. [3, 4]), the pore size within the GDL is locally increased either by laser or mechanical perforation. The "local coating" approach considers the application of a hydrophobic coating to defined regions, leaving the remaining carbon fibers uncoated [5].

Utaka et al. applied locally the hydrophobic coating creating a patterned gas diffusion layer and showed improved oxygen diffusivity using an ex-situ experiment [6]. In a follow-up work the authors combined a hybrid GDL with a micro-grooved flow field channel and showed improved fuel cell performance, mainly motivated by the flow field channels [7].

A recently proposed approach [8–11] considers the use of a novel type of GDL. The new GDL design is a succession of hydrophobic and hydrophilic regions, which are produced in a two-step process. In the first step, the material is modified by using the radiation grafting method which is based on electron irradiation. Steel masks are used to irradiate only the desired regions of the GDL. In a second step, a polymerization reaction is used to produce the hydrophilic surfaces on the desired regions. The hydrophilic regions are expected to liberate the hydrophobic domains from liquid water, therefore increasing the local oxygen diffusivity. By an ex-situ capillary pressure experiment, it was shown that the water preferentially fills the hydrophilic domains while significantly higher pressures are needed in order to fill the hydrophobic areas [11, 12]. Furthermore, cells containing the modified GDLs showed improved performances at elevated current densities in operando. However, finding an optimal pattern designs from experimentation alone can be challenging and time consuming. In such a case, modeling can be used as a powerful tool to study the behavior of the new GDL material.

To model the influence of the succession of hydrophobic and hydrophilic regions, the water transport and the saturation in the GDL has to be described appropriately. Several approaches to model the water transport in GDLs have been reported in literature. A recent state-of-the-art review is given in [13]. The most simple approach is to assume that the liquid water is transported in a vapor phase by the means of Fickean diffusion. The two-phase model approaches treat the liquid water with a separate Darcy equation. The permeability of liquid water and the effective diffusivity of water vapor are then formulated as a function of the saturation. Different models are used to describe the relations between the capillary pressure, saturation and the relative permeabilities. The most common are the Leverett J-function [14], the van Genuchten model [15] and the Brooks-Corey model [16]. The multiphase mixture approach [17] assumes an equilibrium between the gas and liquid phase. It considers that the liquid and vapor phases move simultaneously but with different velocities.

*Corresponding Author: juergen.schumacher@zhaw.ch

A detailed study of the two-phase flow of water reveals that the liquid water transport and the vapor transport are not only coupled through kinematic equations and transport properties, but are also subjected to the heat transport and the phase-change-induced flow [18, 19], also known as the heat-pipe effect. The water is transported along the gradient in the water vapor-pressure (related to the gradient in the temperature distribution) and it condenses around the gas channel. At elevated temperatures, this transport mechanism can even be dominant for the water transport from the GDL into the gas channel.

In a recent work Takaya and Araki [20] developed a three-dimensional numerical model to evaluate the effectiveness of the hybrid GDLs to reduce water saturation under flooding conditions.

In this work we present a numerical model of PEMFCs with position dependent GDL properties to represent the hydrophilic and the hydrophobic regions. The aim of our work is to study the patterned wettability effects and to bring the numerical model one step closer towards the predictive level, which will be used to find the optimal pattern design. The main features of the model are: (i) the structural mechanics part representing the mechanical assembly procedure, (ii) the van Genuchten based two-phase flow, (iii) the condensation and the evaporation of water, (iv) the convective and diffusive transport of the gas species and (v) the electrochemical reactions. The present model is isothermal and the phase-change-induced flow effects are neglected due to the relatively low temperature of the fuel cell.

The paper is organized as follows. In Section 2 the details of the model are presented and in Section 3 the experimental and numerical simulation results are given. The capillary pressure imbibition experiment is presented in Section 3.1, the same problem is examined by using the numerical model in Section 3.2, the simulation results of the in-situ operating cell are presented in Section 3.3 and the results of a parametric study are shown in Section 3.4. Finally, in Section 4 some conclusions are drawn and an outlook is given.

## 2 Numerical model

In this section a three-dimensional numerical model of PEMFC is presented. The geometry of the present model is shown in Fig. 1. The focus of the model is the cathode side GDL which is subdivided into regions of hydrophilic and hydrophobic material. The influences of fuel cell components adjacent to the GDL are introduced into the model as boundary conditions. At the bottom of the GDL an interface with the catalyst layer (CL) is considered. Two boundary condition domains are considered at the top of the GDL: (i) the domains under the ribs and (ii) the domains under the channels.

The model is implemented in COMSOL Multiphysics which includes the following features: the mechanical model simulating the influence of assembly procedure on the porous material properties, the two-phase mass transfer model including condensation/evaporation, the transport of gas species, and the electrochemical interface representing the oxygen reduction reaction.

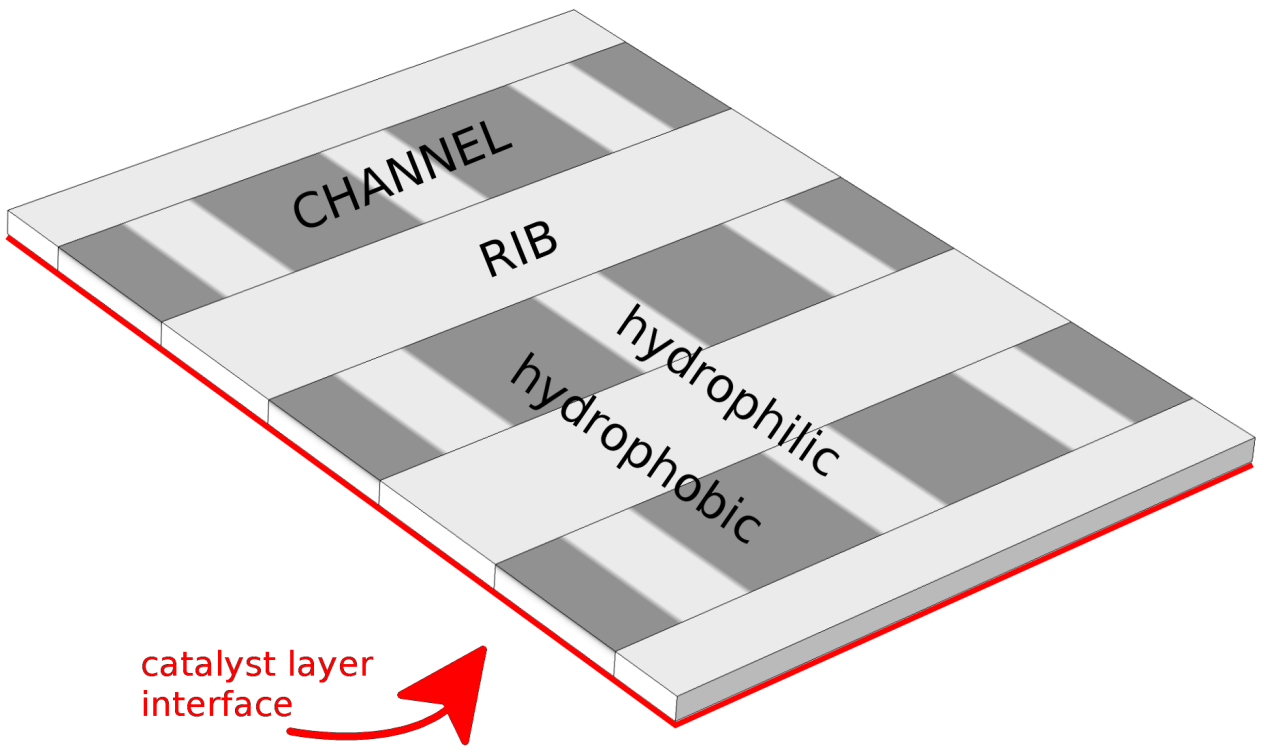

Figure 1: Geometry of the fuel cell model with hydrophilic/hydrophobic pattern of the cathode side GDL. A bipolar plate is positioned on the top of the GDL and the CL is at the bottom. Corresponding boundary conditions to the channel and the rib regions are defined at the top. Channels/ribs are perpendicular to the GDL pattern. At the bottom, a boundary catalyst layer interface is assumed.

*Mechanical model.* – When a fuel cell is assembled, the bipolar plates (BPPs) compress the membrane electrode assembly (MEA). This MEA compression has a significant influence on the fuel cell's performance, since mechanical deformation changes the intrinsic (undeformed) material properties of the MEA components. This shows mainly as the change in porosity of the material, since the compression mostly causes the pore space of the GDL to reduce.

By assuming that the solid material is not compressed and that the compression is only applied in the through-plane direction, one can easily estimate the compressed porosity $\epsilon_c$ as [21]

$$\epsilon_{\mathrm{c}} = 1 - \frac{\delta_0}{\delta_{\mathrm{c}}}(1-\epsilon_0), \tag{1}$$

where $\epsilon_0$ is the porosity of the uncompressed material and $\delta_0$ and $\delta_c$ are the initial and compressed thicknesses, respectively. A step further, what we propose in this paper, is to include the simulation of the mechanical deformation related to the assembly procedure and by using its results determine the spatially resolved values of the porosity in the compressed state. For that purpose we first solve the set of three equations describing the deformation of the linearly elastic body

$$-\nabla\cdot\boldsymbol{\sigma} = \mathbf{0}, \quad \boldsymbol{\varepsilon} = \frac{1}{2}\left((\nabla\boldsymbol{u})^{\mathrm{T}} + \nabla\boldsymbol{u}\right), \quad \boldsymbol{\sigma} = \boldsymbol{C}\boldsymbol{\varepsilon} \tag{2}$$

where $\boldsymbol{\sigma}$ is the stress tensor, $\boldsymbol{\varepsilon}$ is the strain tensor related to the displacement vector $\boldsymbol{u}$ and $\boldsymbol{C}$ is the constitutive matrix. Here, we assume an isotropic material response, therefore the constitutive matrix $\boldsymbol{C}$ is dependent only on the Young's modulus $E$ and Poisson's ratio $\nu$.

In our model we assume that all volumetric change that is caused by the mechanical compression is related to closing of the voids. Therefore we relate the volumetric strain

$$\varepsilon_{\mathrm{v}} = \mathrm{tr}\left(\boldsymbol{\varepsilon}\right), \tag{3}$$

with the spatially distributed value of the effective porosity

$$\epsilon_{\mathrm{eff}} = \frac{\epsilon_0 + \varepsilon_{\mathrm{v}}}{1+\varepsilon_{\mathrm{v}}}, \tag{4}$$

where we considered Eq. (1) and replaced the unidirectional compression term with the generalized term representing the change of volume ($\delta_0/\delta_c \Rightarrow V_0/V = 1/(1+\epsilon_v)$).

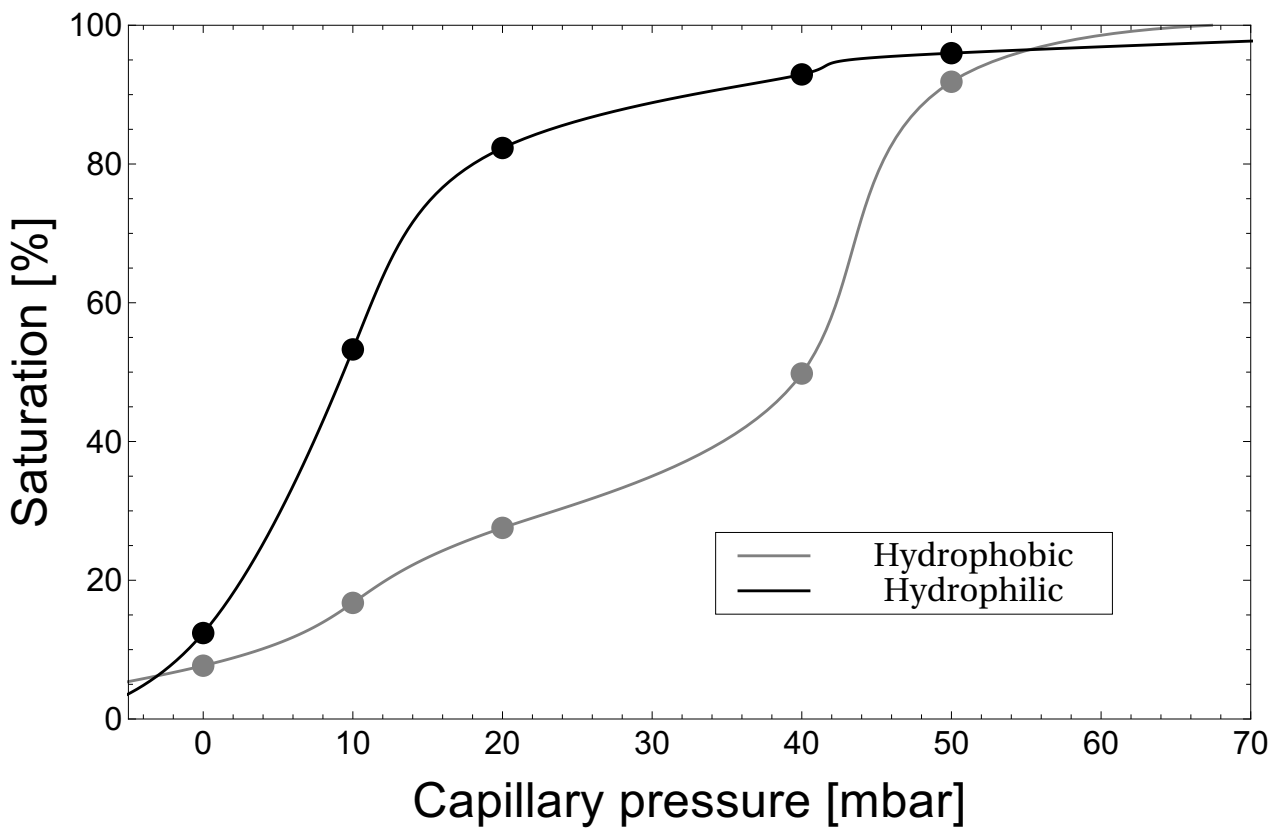

Figure 2: Capillary pressure-saturation curves of the hydrophobic and the hydrophilic regions. The two curves are constructed from the experimental results given in [11]. The experimental data set has been divided into hydrophilic and hydrophobic regions. For each region the average value of the saturation was determined at every value of the applied capillary pressure (black and gray dots on the curves). The final form of each curve is determined by a piece-wise cubic interpolation of the obtained averaged values.

Note that there is only a one-way coupling between the mechanical part of the simulation and the rest of the simulation. We can therefore simulate the compression first and in the subsequent step use the results of the effective porosity $\epsilon_{\text{eff}}$ in the mass transport simulation.

*Two-phase transport of water.* – We assume that the two-phase flow in the porous media is determined by two separate Darcy equations for the gas (gas) and the liquid (liq) phase:

$$\nabla \cdot (\rho_{\text{gas}} \boldsymbol{u}_{\text{gas}}) = Q_{\text{gas}}, \quad \boldsymbol{u}_{\text{gas}} = -\frac{K_{\text{abs}} K_{\text{rel,gas}}}{\mu_{\text{gas}}} \nabla p_{\text{gas}}, \quad (5)$$

$$\nabla \cdot (\rho_{\text{liq}} \boldsymbol{u}_{\text{liq}}) = Q_{\text{liq}}, \quad \boldsymbol{u}_{\text{liq}} = -\frac{K_{\text{abs}} K_{\text{rel,liq}}}{\mu_{\text{liq}}} \nabla p_{\text{liq}}, \quad (6)$$

where $\rho_{\text{gas}}$ and $\rho_{\text{liq}}$ are the densities, $\boldsymbol{u}_{\text{gas}}$ and $\boldsymbol{u}_{\text{liq}}$ are the velocities, $K_{\text{abs}}$ is the intrinsic material permeability, $K_{\text{rel,gas}}$ and $K_{\text{rel,liq}}$ are the relative permeabilities, $\mu_{\text{gas}}$ and $\mu_{\text{liq}}$ are the viscosities, $p_{\text{gas}}$ and $p_{\text{liq}}$ are the pressures and $Q_{\text{gas}}$ and $Q_{\text{liq}}$ are the mass source/sink terms. Note that in fuel cell simulations, when dealing with only non-modified GDL material, the gas phase plays the role of the wetting phase while the water plays the role of the non-wetting phase. This is also true for the hydrophobic regions in our case while the roles of the phases are reversed in the hydrophilic regions.

The saturation part of the model is based on the Van Genuchten model, which is expressed in terms of the capillary pressure $p_{\text{c}} = p_{\text{liq}} - p_{\text{gas}}$. The saturation of the liquid phase is given by [22]

$$s\,(p_{\text{c}}) = \left(1 + \left(\frac{1\,\text{atm} - p_{\text{c}}}{p_{\text{b}}}\right)^{m}\right)^{-n}, \quad (7)$$

where $p_{\text{b}}$, $m$ and $n$ are material dependent parameters. Instead of finding the best fit for parameters $p_{\text{b}}$, $m$ and $n$ we plugged the experimentally obtained saturation versus the capillary pressure curves directly into our model. The saturation curves for the hydrophilic and hydrophobic domains are presented in Fig. 2.

The relative permeabilities of the two phases read [23]

$$K_{\text{rel,gas}}\,(s) = s^{k}\left(1 - \left(1 - s^{1/m}\right)^{m}\right)^{2}, \quad (8)$$

$$K_{\text{rel,gas}}\,(s) = (1-s)^{k}\left(1 - s^{1/m}\right)^{2m}, \quad (9)$$

where $k$ and $m$ are fit parameters. (hydrophobic)

*Phase change.* – The volumetric rate of interfacial mass transfer between the liquid phase and the vapor phase of water during evaporation and condensation is modeled by using the Hertz-Knudsen-Langmuir equation [24] as

$$Q_{\text{pc}} = \begin{cases} \frac{A_{\text{pore}}}{M_{\text{w}}} \text{Sh}_{\text{c}} \frac{D_{\text{v}}}{\overline{d}} (\rho_{\text{v}} - \rho_{\text{sat}})(1-s) & \text{if } \rho_{\text{v}} \geq \rho_{\text{sat}} \\ \frac{A_{\text{pore}}}{M_{\text{w}}} \text{Sh}_{\text{e}} \frac{D_{\text{v}}}{\overline{d}} (\rho_{\text{v}} - \rho_{\text{sat}}) s & \text{if } \rho_{\text{v}} < \rho_{\text{sat}} \end{cases}, \quad (10)$$

where $A_{\text{pore}}$ is the pore surface area per unit volume, $\rho_{\text{v}}$ is the water vapor mass density, $\rho_{\text{sat}}$ is the mass density of saturated water vapor, $M_{\text{w}}$ is the molar mass of water, $\text{Sh}_{\text{c,e}}$ are Sherwood numbers accounting for the mass transport capability during condensation/evaporation, given by

$$\text{Sh}_{\text{c,e}} = \Gamma_{\text{s}}\,\Gamma_{\text{m}} \sqrt{\frac{RT}{2\pi M_{\text{w}}}}\,\frac{\overline{d}}{D_{\text{v}}}, \quad (11)$$

where $\overline{d}$ is the characteristic length for water diffusion and $D_{\text{v}}$ is the mass diffusivity of water vapor, which both cancel out. $\Gamma_{\text{m}}$ is an uptake coefficient that accounts for the combined effects of heat and mass transport limitations in the vicinity of the liquid/vapor interface, and $\Gamma_{\text{s}}$ is the interfacial area accommodation coefficient [24]. By using Eq. (10) we determine the mass source/sink terms for Eqs. (5) and (6):

$$Q_{\text{gas}} = -Q_{\text{pc}} M_{\text{w}}, \quad Q_{\text{liq}} = Q_{\text{pc}} M_{\text{w}}. \quad (12)$$

*Transport of species.* – We consider the concentrations of oxygen $c_{\text{O}_2}$ and water vapor $c_{\text{v}}$ as the variables of the problem. By considering both the convective and the diffusive term, the species transport equation is

$$\nabla \cdot j_{\alpha} = \nabla \cdot (\underbrace{-D_{\alpha} \nabla c_{\alpha}}_{j_{\alpha,\text{diff}}} + \underbrace{c_{\alpha} \boldsymbol{u}_{\text{gas}}}_{j_{\alpha,\text{conv}}}) = Q_{\alpha}, \quad \alpha = \text{O}_2, \text{v} \quad (13)$$

where $D_{\alpha}$ is the diffusion coefficient, the velocity field $\boldsymbol{u}_{\text{gas}}$ is obtained from the Darcy's flow solution of Eq. (5), $j_{\alpha}$ is the total flux of species $\alpha$, $j_{\alpha,\text{diff}}$ is the diffusive part of the flux, $j_{\alpha,\text{conv}}$ is the convective part of the flux, $Q_{\alpha}$ is the species sink/source term and index $\alpha$ denotes one of the species. The oxygen sink term is equal to zero ($Q_{\text{O}_2} = 0$), since there are no chemical reactions present in the GDL. The water vapor sink/source term is set to

$$Q_{\text{v}} = -Q_{\text{pc}}, \quad (14)$$

to take into account the condensation and evaporation of water.

According to Chapman-Enskog theory, the intrinsic binary diffusion coefficients of water vapor and oxygen in oxygen can be approximated by [25]

$$D_{\text{v}}^{\text{int}} = 2.77 \times 10^{-5} \left[\frac{\text{m}^2}{\text{s}}\right] \left(\frac{T}{T_{\text{ref}}}\right)^{3/2} \frac{p_{\text{ref}}}{p_{\text{gas}}} \quad (15)$$

and

$$D_{\text{O}_2}^{\text{int}} = 2.06 \times 10^{-5} \left[\frac{\text{m}^2}{\text{s}}\right] \left(\frac{T}{T_{\text{ref}}}\right)^{3/2} \frac{p_{\text{ref}}}{p_{\text{gas}}} \quad (16)$$

where $T_{\text{ref}} = 25°\text{C}$ and $p_{\text{ref}} = 1\,\text{atm}$ are the reference consitions.

In Eq. (13) we also consider the effects of liquid water on the gas transport by calculating the effective diffusion coefficients

$$D_{\alpha} = D_{\alpha}^{\text{int}} \left(\frac{(1-s)\epsilon_{\text{eff}}}{\tau}\right)^{1.5}. \quad (17)$$

The parameterization in Eq. (17) is similar to the one proposed in Um and Wang [26] with the difference that we also include the tortuosity $\tau$ and we do not consider the geometry factor to account for the shadowing effect of the ribs, since this is already accounted for with 3D modeling and the boundary conditions.

*Electrochemical interface.* – At the boundary between the GDL and the CL we assume an electrochemical interface, where oxygen is consumed and water is produced. In this simplified model, we only consider the cathode half-cell reaction. The production/consumption rates are dependent on the local current density $i_{\mathrm{loc}}$, which is given by the Butler-Volmer equation [27–29]

$$i_{\mathrm{loc}} = i_0 \left( \exp\left[ \frac{2\alpha_{\mathrm{C}} F}{RT} \eta_{\mathrm{C}} \right] - \exp\left[ -\frac{2(1-\alpha_{\mathrm{C}})F}{RT} \eta_{\mathrm{C}} \right] \right), \tag{18}$$

where $F$ is the Faraday constant, $R$ is the gas constant, and $i_0$ is the exchange current density of the cathode at standard conditions. We express the oxygen reduction reaction overpotential $\eta_{\mathrm{C}}$ as

$$\eta_{\mathrm{C}} = E_{\mathrm{eq}} - E_{\mathrm{cell}} - \eta_{\mathrm{R}} - \eta_{\mathrm{OC}}, \tag{19}$$

where $\eta_{\mathrm{R}}$ and $\eta_{\mathrm{OC}}$ are overpotential terms detailed further below, $E_{\mathrm{cell}}$ denotes the cell voltage, and the equilibrium potential is given by the Nernst equation

$$E_{\mathrm{eq}} = -\frac{\Delta G}{2F} + \frac{RT}{2F} \ln\left[ \left( \frac{p_{\mathrm{O_2}}}{p_{\mathrm{ref}}} \right)^{1/2} \right], \tag{20}$$

in which $p_{\mathrm{ref}} = 1\,\mathrm{atm}$ is the standard atmospheric pressure and $p_{\mathrm{O_2}}$ denotes the partial pressure of oxygen. For simplicity, we assumed here that the oxygen activity is the rate-limiting factor, such that the electron, proton and water activities, that appear in the full generalized Butler-Volmer equation [29], can all bet set to unity. $\Delta G$ denotes the Gibbs free energy of the reaction at operating conditions, assuming temperature-indendence of the reaction entropy $\Delta S$, which holds in very good approximation [28]:

$$\Delta G = \Delta G_{\mathrm{ref}} - (T - T_{\mathrm{ref}}) \Delta S_{\mathrm{ref}} \tag{21}$$

The corresponding reference values $\Delta G_{\mathrm{ref}}$ and $\Delta S_{\mathrm{ref}}$ at reference temperature $T_{\mathrm{ref}}$ and pressure $p_{\mathrm{ref}}$ can be found in the standard thermodynamics literature [30] and are given in the parameter list at the end of the article.

To account for voltage losses from ohmic resistance of the catalyst-coated membrane (CCM), which is not spatially resolved in the present model, we use an ohmic overpotential term $\eta_{\mathrm{R}}$ in Eq. (19), given by

$$\eta_{\mathrm{R}} = i_{\mathrm{loc}} \frac{d_{\mathrm{CCM}}}{\sigma_{\mathrm{p}}}, \tag{22}$$

where $d_{\mathrm{CCM}} = d_{\mathrm{m}} + 2d_{\mathrm{CL}}$ is the thickness of the CCM, whereas $\sigma_{\mathrm{p}}$ is its protonic conductivity, given by [31]

$$\sigma_{\mathrm{p}} = \epsilon_{\mathrm{i}}^{1.5} (0.514\lambda - 0.326) \exp\left[ 1268\,\mathrm{K} \left( \frac{1}{303\,\mathrm{K}} - \frac{1}{T} \right) \right], \tag{23}$$

which is dependent on temperature $T$, the ionomer volume fraction $\epsilon_{\mathrm{i}}$, and the level of the water content in the ionomer $\lambda$, given by the ratio of water molecules per sulfonic acid group $H_2O/SO_3^-$. In order to condense the proton-conducting properties of the whole CCM into this zero-dimensional expression, we average the ionomer volume fraction over the whole CCM, resulting in an effective (averaged) value

$$\epsilon_{\mathrm{i}} = \frac{d_{\mathrm{m}} \epsilon_{\mathrm{i,m}} + 2 d_{\mathrm{CL}} \epsilon_{\mathrm{i,CL}}}{d_{\mathrm{CCM}}} < 1, \tag{24}$$

hence the Bruggeman correction in Eq. (23). All other activation losses (such as those resulting from gas crossover, for instance) are combined in the constant open circuit overpotential $\eta_{\mathrm{OC}}$.

*Transport of dissolved water.* – The transport of dissolved water in the ionomer is modeled according to Springer at al. [32]. The net flux of dissolved water is composed of two additive terms describing electroosmotic drag and the back diffusion of water:

$$j_\lambda = 2n_{\mathrm{drag}}\, j_{\mathrm{H_2}} - \frac{D_\lambda}{V_{\mathrm{m}}} \nabla\lambda, \quad n_{\mathrm{drag}} = \frac{2.5\lambda}{22}, \tag{25}$$

where $n_{\mathrm{drag}}$ is the electroosmotic drag coefficient, $j_{\mathrm{H_2}} = i_{\mathrm{loc}}/2F$ is the molar flux of hydrogen (half of the flux of the hydrogen ions), $D_\lambda$ is the diffusion coefficient and $V_{\mathrm{m}} = M_{\mathrm{m}}/\rho_{\mathrm{dry}}$ is the molar volume of the dry membrane. The diffusion coefficient $D_\lambda$ can be parameterized as [33]

$$D_\lambda = \begin{cases} \left(7.32 \times 10^{-8} \exp[0.12\lambda] + 5.41 \times 10^{-10} \exp[1.44\lambda]\right) \\ \quad \times \exp[-2436\,\mathrm{K}/T] \quad \text{for } \lambda < 4 \\ \left(1.58 \times 10^{1} \exp[-4.66\lambda] + 1.45 \times 10^{-7} \exp[0.04\lambda]\right) \\ \quad \times \exp[-2436\,\mathrm{K}/T] \quad \text{for } \lambda \geq 4 \end{cases} \tag{26}$$

The following averaged gradient is assumed to evaluate the diffusion contribution in Eq. (25):

$$\nabla\lambda = \frac{\lambda_{\mathrm{C}} - \lambda_{\mathrm{A}}}{d_{\mathrm{CCM}}}, \tag{27}$$

where $\lambda_{\mathrm{A}}$ and $\lambda_{\mathrm{C}}$ are the water content values evaluated at the anode and the cathode side, respectively. The drag contribution in Eq. (25) is determined by assuming the average membrane water content $\overline{\lambda} = (\lambda_{\mathrm{A}} + \lambda_{\mathrm{C}})/2$. The relationship between the relative humidity RH and the equilibrium membrane water content is given by

$$\lambda^{\mathrm{eq}} = \begin{cases} 0.043 + 17.81\mathrm{RH} - 39.85\mathrm{RH}^2 + 36\mathrm{RH}^3, & \mathrm{RH} \leq 1 \\ \lambda^{\mathrm{eq}}(\mathrm{RH}=1) + \frac{22 - \lambda^{\mathrm{eq}}(\mathrm{RH}=1)}{2} (\mathrm{RH} - 1), & \mathrm{RH} > 1 \end{cases}, \tag{28}$$

where the polynomial fit of [32] is considered for relative humidities below 1 and a linear extrapolation to the maximum $\lambda = 22$ is assumed for higher humidities.

The local current density is treated as a dependent variable of the problem and we obtain its spatial distribution by solving Eqs. (18) to (28). With $i_{\mathrm{loc}}$ defined we can determine the oxygen and the water vapor boundary fluxes at the GDL and the CL interface

$$\overline{Q}_{\mathrm{O_2}} = -\frac{i_{\mathrm{loc}}}{4F}, \quad \overline{Q}_{\mathrm{v}} = \frac{(1 + 2n_{\mathrm{drag}}) i_{\mathrm{loc}}}{2F} - \frac{D_\lambda}{V_{\mathrm{m}}} \nabla\lambda, \tag{29}$$

where $\overline{Q}_{\mathrm{v}}$ contains the contribution of the chemically produced water as well as the contribution of water transport through the membrane according to Eq. (25).

*Solution method.* – The model was implemented in COMSOL Multiphysics which uses the finite element method (FEM) to find the numerical solution to the boundary value problem (see e.g. [34–36]). The basic principle of the FEM considers mesh discretization of a continuous domain into a set of discrete sub-domains, called elements. The finite element meshes used to obtain the results in Sections 3.2 and 3.3 are presented in Figs. 4 and 7.

Further, instead of directly solving the partial differential equations (PDEs), FEM first converts the original strong form PDEs into their equivalent weak formulations. In the present situation, a steady state case, the strong form PDEs (2), (5), (6) and (13) are therefore locally (within each element of the mesh) approximated with a set of algebraic equations dependent on the nodal degrees of freedom (DOFs). The global system of equations is obtained by systematically recombining all sets of the element equations into a global system of equations, which uses a Newton iteration procedure to obtain the numerical solution.

In the present model we used quadratic shape functions to represent the displacement field $\boldsymbol{u}$, the pressure $p_{\mathrm{liq}}$ and the pressure $p_{\mathrm{gas}}$, while linear shape functions were employed for the concentrations $c_{\mathrm{O}_2}$ and $c_{\mathrm{v}}$ and also to represent the variable related to local current density $i_{\mathrm{loc}}$. In combination with the chosen mesh presented in Fig. 7, this resulted in 591705 global DOFs to simulate the operating fuel cell in Section 3.3. Note that the computation effort is significantly reduced when one-way coupling between the displacement field and the rest of the fields is considered. Hence, in the first step we only considered 319599 DOFs related to $\boldsymbol{u}$ and in the subsequent step(s) we only considered 272106 DOFs related to $p_{\mathrm{liq}}$, $p_{\mathrm{gas}}$, $c_{\mathrm{O}_2}$, $c_{\mathrm{v}}$, $i_{\mathrm{loc}}$.

## 3 Results

In this section we present the results obtained experimentally and also the results obtained by using numerical simulation. First, we show the results of the ex-situ capillary pressure experiment. Next, we present the results of the ex-situ capillary pressure simulation and compare them to the experimental data. Further, the simulation results of the in-situ operating cell are presented. Last, the results of a parametric study are given.

### 3.1 Ex-situ capillary pressure experiment

A Toray TPG-H-060 (purchased at Fuel Cell Earth) GDL was used as substrate. The plain GDLs were in-house coated with fluoroethylenpropylene (FEP) at 30% following a dip coating procedure [11]. The materials were then irradiated using 2 mm thick (with 500 $\mu$m wide rectangular slits regularly spaced 950 $\mu$m apart) steel masks with 200 keV electrons (EBLab 200, Comet AG) receiving a dose of 50 kGy. The materials were then reacted with pure N-vinylformamide (purchased at Sigma Aldrich) during 60 minutes at 70°C in oxygen free atmosphere. The capillary pressure experiments were performed using the setup detailed in [11]. A cell with 1×1 cm of active area was used with a bipolar plate containing a single water injection channel (1 mm width) in the center. The modified material was positioned between the bipolar plate and a hydrophobic membrane (HVHP04700, Durapore), which allows having the gas equilibrated at atmospheric pressure and it is meanwhile impermeable for liquid water. A regular (unmodified) GDL was used in the other half of the cell and the GDL compression was set to 20%. The experiment was carried out at 25°C. The liquid pressure was controlled by regulating the height of a water tank controlled by a precision motor and was changed in increasing fashion. Neutron images were recorded during the whole duration of the experiments. Details about neutron radiography and image processing can be found in [11, 37, 38].

In Fig. 3 the total water thickness distribution obtained by neutron radiography imaging is presented. The line with the maximum value of water thickness spanning from far left to far right represents the water injection channel, which was also captured by the neutron imaging. The modified regions of material (hydrophilic zones) are the narrow strips bounded by the dashed borders. It is evident that water accumulates in these regions and that the water distribution in these zones is fairly uniform. Note that the water is not strictly confined to the hydrophilic zones thus resulting in a slightly wider strip of water than the width of the modified material.

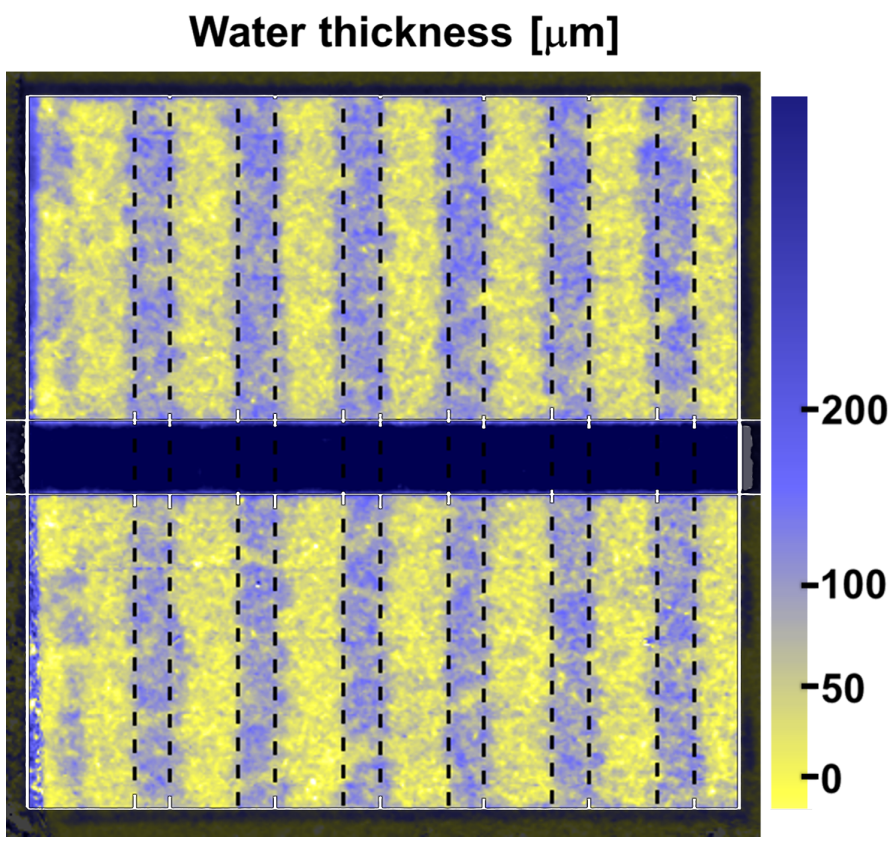

Figure 3: Experimentally obtained distribution of the total water thickness in the patterned GDL at a capillary pressure of 3 kPa. Neutron imaging also captured the water in the injection channel spanning from left to right. The water accumulates in the hydrophilic domains. The width of the water strip is slightly wider than the width of the modified material.

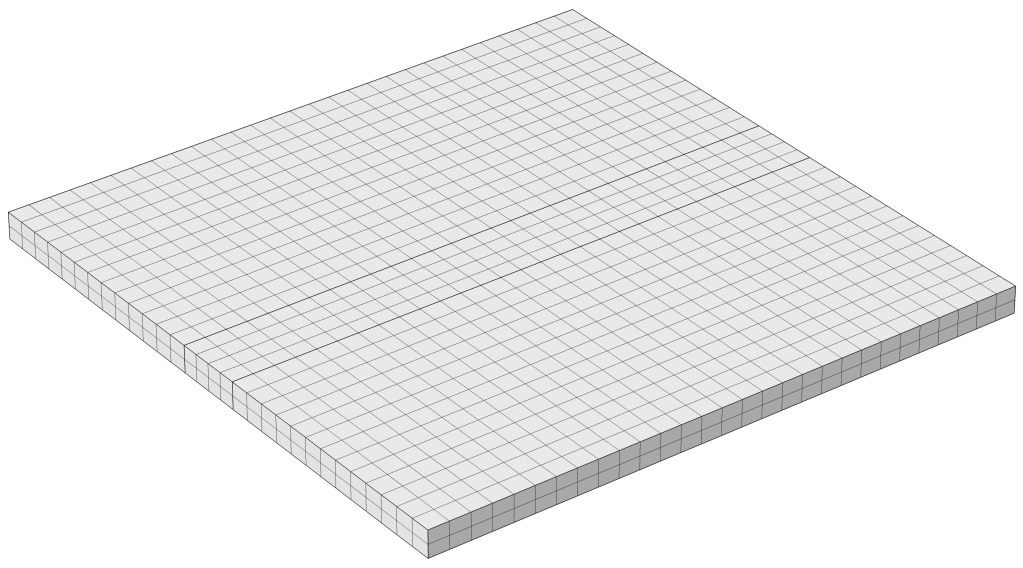

Figure 4: Finite element mesh used in the ex-situ capillary pressure simulation. The mesh consists of 1740 cuboid finite elements. 3 × 17995 DOFs are used to represent the displacement field $\boldsymbol{u}$ and 2 × 17995 DOFs are used to represent the pressures $p_{\mathrm{liq}}$ and $p_{\mathrm{gas}}$.

### 3.2 Ex-situ capillary pressure simulation

The problem experimentally addressed in the previous section is now examined by using the numerical model. Note that in this example we are only using the mechanical part and the two-phase flow part of the model presented in Section 2. The region of interest in our simulation is an isolated GDL which is 190 $\mu$m thick and $L = 4350\,\mu$m long in both $X$ (perpendicular to the pattern) and $Y$ direction. The GDL pattern consists of 500 $\mu$m wide hydrophilic regions and 950 $\mu$m wide hydrophobic regions, thus the area of interest covers 3 hydrophilic and 3 hydrophobic regions. Note that in simulations we also consider a transitional region (225 $\mu$m into the hydrophobic region), where the response of the material is a superposition of the hydrophilic and hydrophobic case. These regions are introduced to simulate the gradual change from one material behavior to the other, which is in agreement with results presented in [11] and also shown in Fig. 3. The compression simulation, by considering $E = 17.9$ MPa and $\nu = 0.3$, was carried out by applying a uniform displacement $u_Z = 38\,\mu$m in the $Z$ direction at the bottom of the GDL, while at the top of the GDL the displacement in $Z$ direction was restrained ($u_Z = 0$). Additional restrains were considered at the planes corresponding to $Y = 0$ and $Y = L$ ($u_Y = 0$) and $X = 0$ and $X = L$ ($u_X = 0$). The water injection was simulated by setting gas pressure at the top of the GDL to the atmospheric pressure ($p_{\mathrm{gas}}^{\mathrm{BC}} = 1\,\mathrm{atm}$) and by varying the liquid water pressure at the 1 mm wide strip at the bottom of the GDL ($p_{\mathrm{liq}}^{\mathrm{BC}}$). In Fig. 4 we plot the mesh used, while the list of parameters used in the simulation is given at the end of the paper.

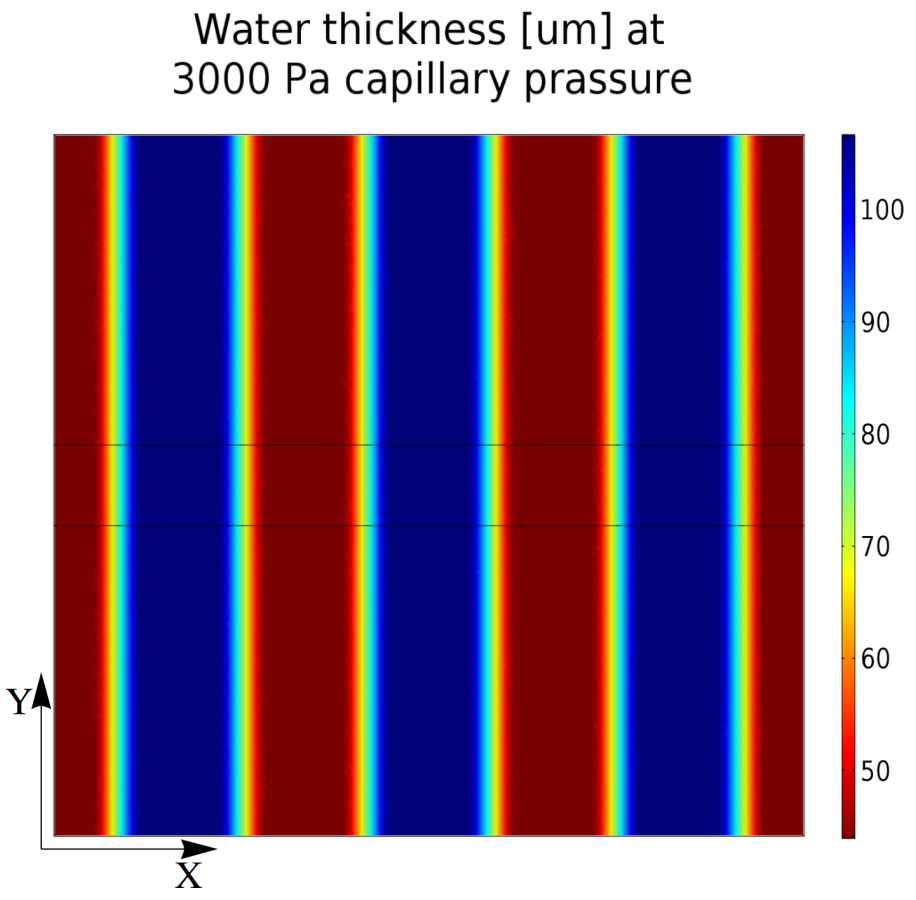

Figure 5: Simulated distribution of the total water thickness in the patterned GDL at a capillary pressure of 3 kPa. The geometry consists of 3 full-sized hydrophilic regions and two full-sized hydrophobic regions confined by two half-sized hydrophobic regions. The water is uniformly distributed in the $Y$ direction. In $X$ direction, a pattern emerges with more water accumulation in the hydrophilic and less water accumulation in hydrophobic regions.

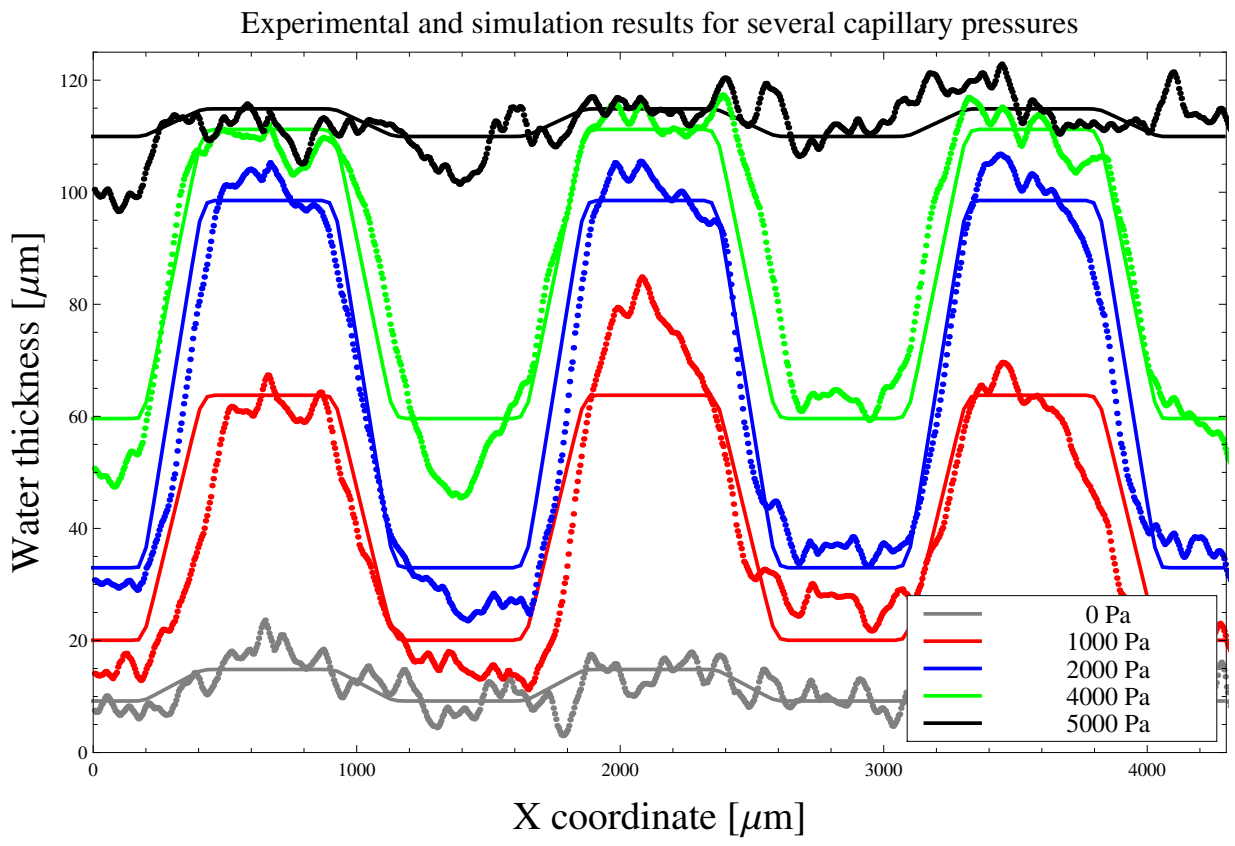

Figure 6: Comparison of experimental (dotted) and simulated (full line) water thicknesses as a function of position for different applied capillary pressures. The succession of hydrophobic and hydrophilic regions is clearly recognizable along the $X$ axis in both the simulated and the experimentally obtained results.

In Figs. 5-6 the results of the simulation are presented. Fig. 5 depicts the distribution of the total water thickness at the capillary pressure equal to $p_{\mathrm{liq}}^{\mathrm{BC}} - p_{\mathrm{gas}}^{\mathrm{BC}} = 3\,\mathrm{kPa}$. The total water thickness is obtained as the integral over the GDL's thickness $\int s\,\epsilon_{\mathrm{eff}}\; dz$. One can see that the water is uniformly distributed in the direction perpendicular to the injection channel. In $X$ direction, an evident pattern emerges with more water accumulation in the hydrophilic (approx. $100\,\mu$m) and less water accumulation in hydrophobic regions (approx. $50\,\mu$m).

In Fig. 6 we compare the simulated water thickness with the experimental results obtained in [11]. The curves correspond to different values of the applied capillary pressure. The results of the present model are in good agreement with the experiment. The succession of hydrophobic and hydrophilic regions is clearly seen along the $X$ axis in both the simulated and the experimentally obtained results.

### 3.3 Operating cell

For the in-situ simulation we considered a region that is $L_X = 2.9\,\mathrm{mm}$ long in the $X$ direction and $L_Y = 4\,\mathrm{mm}$ long in the $Y$ direction. This size of the area includes two full repetitions of hydrophilic/hydrophobic pattern and two repetitions of channel/rib. The repetitions in both directions allow us to significantly reduce the computation time by simulating only the smallest repeating pattern. The mesh shown in Fig. 7 was therefore used to simulate a domain of $L_X/4 \times L_Y/4$ representing one half of hydrophilic and one half of hydrophobic region in $X$ direction and one half of rib and one half of channel region in $Y$ direction. The operating conditions of the base case simulation are shown in Table 1, while the full list of parameters is presented at the end of the paper.

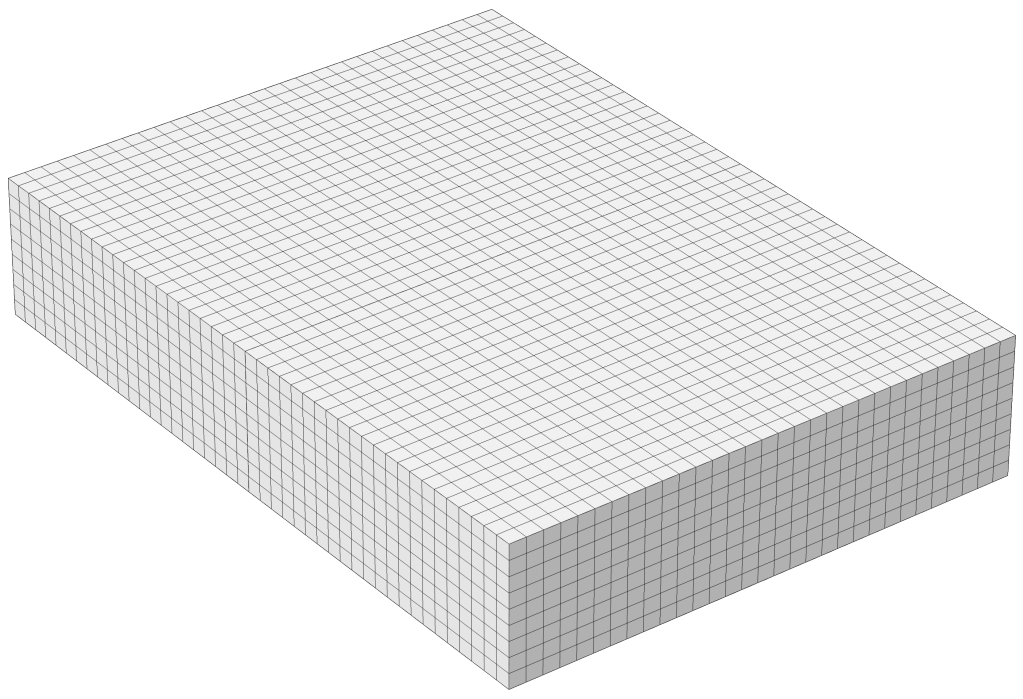

Figure 7: The finite element mesh used in the simulation of the operating cell. Computation time was reduced by considering the pattern repetitions in both $X$ and $Y$ directions thus reducing the simulated domain to $L_X/4 \times L_Y/4$. The mesh consists of 12276 cuboid finite elements. $3 \times 106533$ DOFs are used to represent the displacement field $\boldsymbol{u}$, $2 \times 106533$ DOFs are used to represent the pressures $p_{\mathrm{liq}}$ and $p_{\mathrm{gas}}$, $2 \times 28800$ DOFs are used to represent the concentrations $c_{\mathrm{O}_2}$ and $c_{\mathrm{v}}$, and 1440 DOFs are used to represent the planar distribution of $i_{\mathrm{loc}}$.

The boundary conditions for the GDL mechanical compression simulation are the same as in the ex-situ experiment with the difference that at the bipolar plate side only the regions under the ribs are subjected to imposed displacement $u_Z$ in $Z$ direction.

The boundary conditions for the pressures of the two phases are set in the regions corresponding to the channel floor. By defining the pressures we also, through the definition of the capillary pressure, set the boundary values of saturation. A practice often used in literature [24, 39–42] is to set the saturation boundary condition at the GDL/gas channel interface to the value representing the immobile saturation. Analogous to this approach we set the capillary pressure in the channels region to $p_{\mathrm{c}}^{\mathrm{BC}} = 3.2$ mbar and thus set the saturation level, determined by the curves in Fig. 2, in the hydrophobic regions to 0.100 and in the hydrophilic regions to 0.227. By combining the operating pressure of the simulated fuel cell (2 atm) with the capillary pressure representing the boundary saturations we finally set the phase pressures on the channel floors to $p_{\mathrm{gas}}^{BC} = 2$ atm and $p_{\mathrm{liq}}^{BC} = p_{\mathrm{gas}}^{BC} + p_{\mathrm{c}}^{\mathrm{BC}}$.

Table 1: Operating conditions for the base case simulation.

| | |
|---|---|
| $p_{\mathrm{gas}}^{\mathrm{BC}} = 2\,\mathrm{atm}$ | Boundary pressure of gas phase |
| $p_{\mathrm{liq}}^{\mathrm{BC}} = p_{\mathrm{gas}}^{\mathrm{BC}} + p_{\mathrm{c}}^{\mathrm{BC}}$ | Boundary pressure of liquid phase |
| $p_{\mathrm{c}}^{\mathrm{BC}} = 3.2\,\mathrm{mbar}$ | Boundary capillary pressure |
| $\mathrm{RH_A} = 1.00$ | Relative humidity of anode gases |
| $\mathrm{RH_C} = 0.90$ | Relative humidity of cathode gases |
| $T = 50°\mathrm{C}$ | Temperature of fuel cell |
| $u_Z = -38\,\mu\mathrm{m}$ | Applied BPP displacement |
| $E_{\mathrm{cell}} = 0.6\,\mathrm{V}$ | Cell voltage |
| $\alpha_{\mathrm{O_2}}^{\mathrm{in}} = 0.2$ | Molar fraction of oxygen in channels |

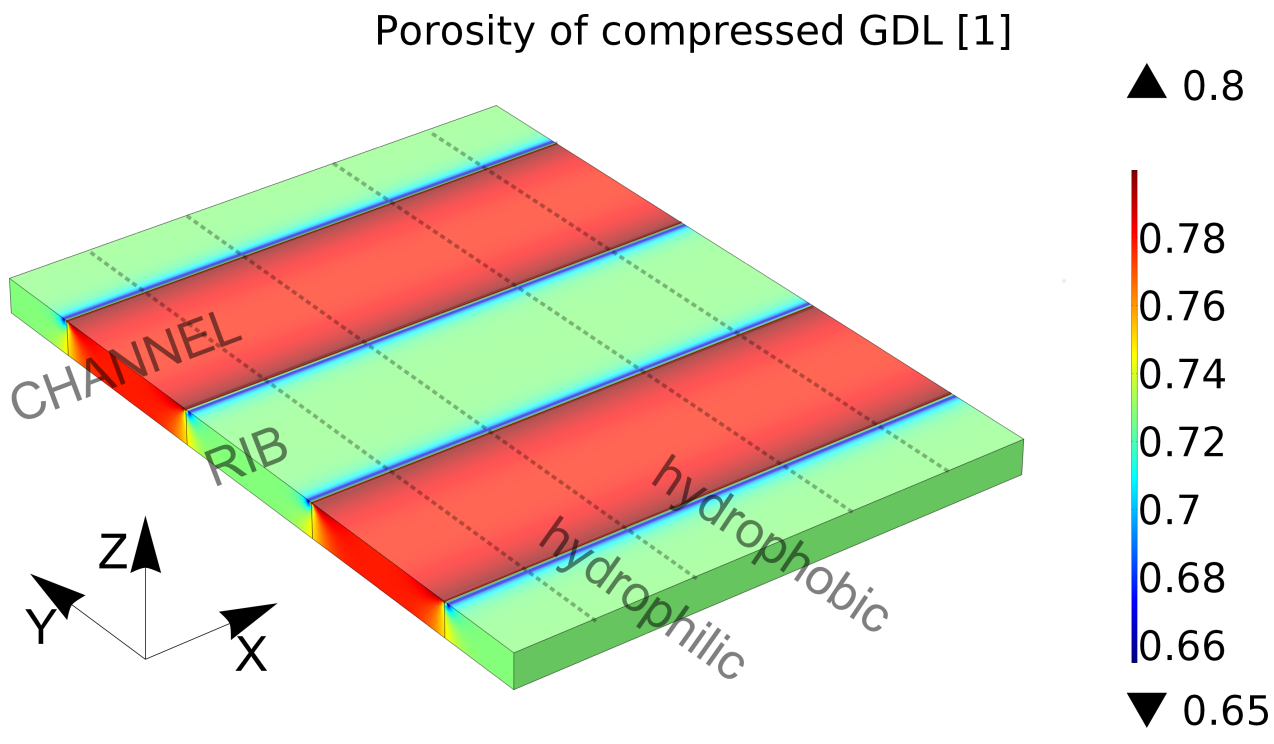

Figure 8: Simulated distribution of the effective porosity of compressed GDL. The values are lower under the ribs due to the higher level of material compression.

The free gas flow in the channel regions is in the present model not considered. Instead we assume constant values of gas species concentrations in the channel floor regions by considering the cathode side relative humidity $\mathrm{RH_C}$ and the molar fraction of oxygen $\alpha_{\mathrm{O_2}}^{\mathrm{in}}$.

The electric operating point of the simulated fuel cell is steered by controlling the fuel cell voltage $E_{\mathrm{cell}}$ in the simulations.

*Porosity of compressed GDL.* – In Fig. 8 we present the distribution of porosity of the compressed GDL. One can see that under the channels we have higher porosity than under the ribs. The material under the ribs is more compressed than the material under the channels. The values under the ribs are around 73% while the porosity under the channels hardly changes and it is around 78%.

*Liquid water.* – In Figs. 9 and 10 we present the results related to the liquid water distribution. In Fig. 9 the capillary pressure distribution is shown. The capillary pressure is the highest under the ribs in the hydrophilic region (3.41 mbar), it is slightly lower under the ribs in the hydrophobic region (3.36 mbar) and the lowest in the channel region (3.20 mbar) where the pressure boundary conditions are set. It has to be mentioned that, although the simulated capillary pressure is within the range of the experimentally obtain capillary-pressure curves in Fig. 2, the simulated variation (0.21 mbar) is much smaller than the difference between two measuring points from the ex-situ experiment (10 mbar). In our simulation, when modeling a very small domain, the conditions resemble those of the differential cell. The in-plane gradients are small since large scale simulation effects, like the transport of liquid water along the gas channels for example, are not considered. Therefore, when dealing with simulation of a larger area one would expect to have greater variation of the capillary pressure. Nevertheless, the quality of simulation results would improve if more than 5 experimental points were available.

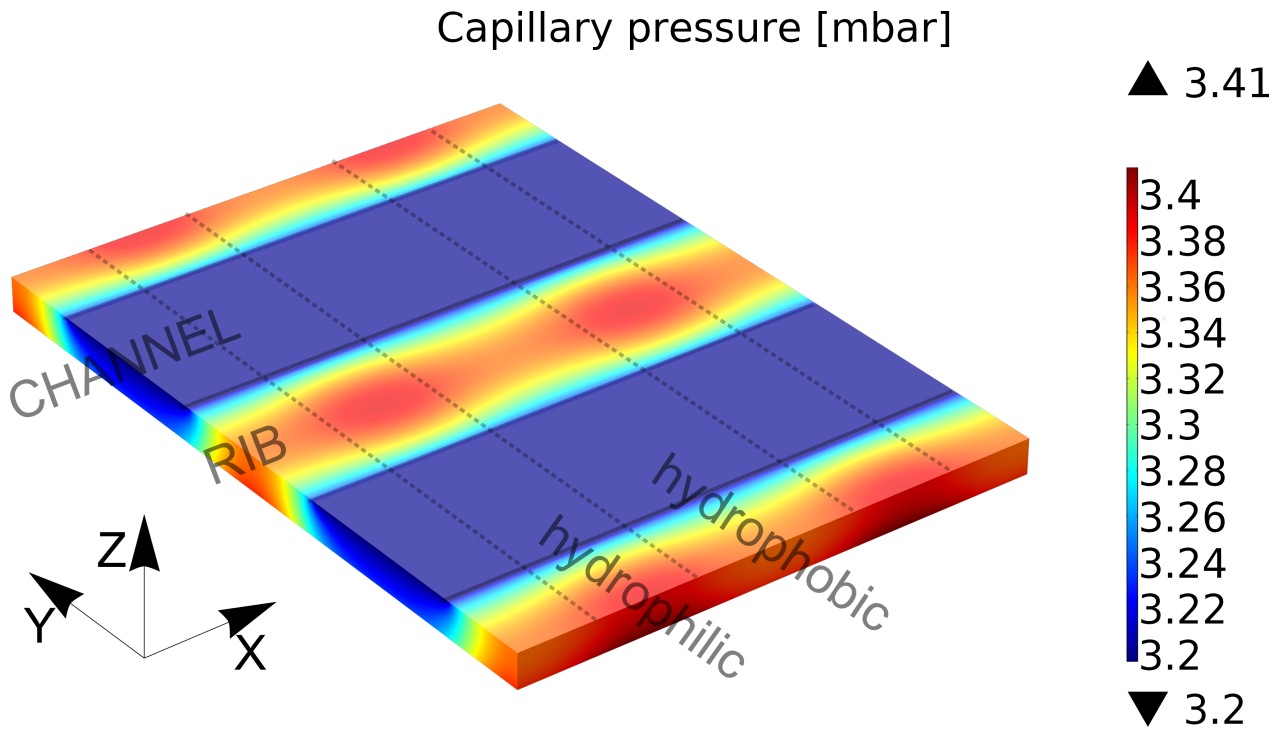

Figure 9: Simulated distribution of the capillary pressure of the patterned GDL at a fuel cell voltage of 0.6 V. The highest values of the capillary pressure are under the ribs in the hydrophilic region (3.41 mbar), slightly lower values are observed under the ribs in the hydrophobic region (3.36 mbar) and the lowest are in the channel region (3.20 mbar) where the boundary conditions are set.

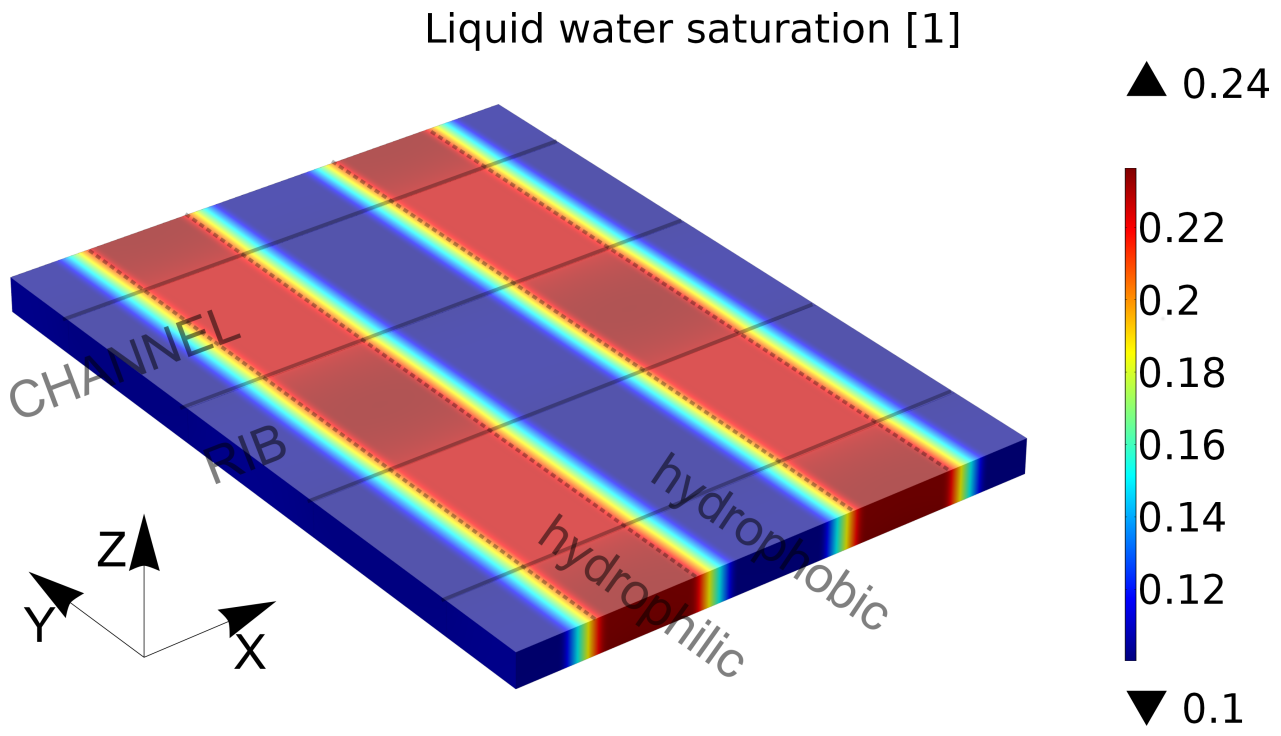

Figure 10: Simulated distribution of the liquid water saturation of the patterned GDL at a fuel cell voltage of 0.6 V. The hydrophilic/hydrophobic pattern is most pronounced in this figure. The highest values (0.236) are on the catalyst side under the ribs in the hydrophilic region. The lowest values (0.100) are observed on the bipolar plate side under the channels in the hydrophobic region.

Fig. 10 depicts the distribution of the liquid water saturation. Here, the hydrophilic/hydrophobic pattern is most pronounced. However, although difficult to observe from Fig. 10, there is a variation between the channels and ribs and there is also a through-plane variation of values. The highest values are observed under the ribs in the hydrophilic regions. The value on the catalyst side is 0.236 while the value on the bipolar plate side is slightly lower at 0.235. The lowest values are observed on the bipolar plate side under the channels in the hydrophobic region (0.100) where the pressure boundary condition is set.

*Phase change rates.* – Figure 11 shows the distribution of the phase change rates. Positive values indicate condensation while negative values indicate evaporation. Here the effect of both channel/rib and the hydrophilic/hydrophobic pattern is evident. We have the lowest rates (evaporation) under the channel regions, which reflect the lower humidity of the inlet gases. In the channel region one can also see the hydrophilic/hydrophobic pattern. This is caused by the different levels of saturation, which also plays a role in the phase change process (see Eq. 10). The condensation takes place under the ribs and in the vicinity of the CL boundary, which is related to higher water vapor concentration caused by the electrochemically produced water and the water dragged from the anode side.

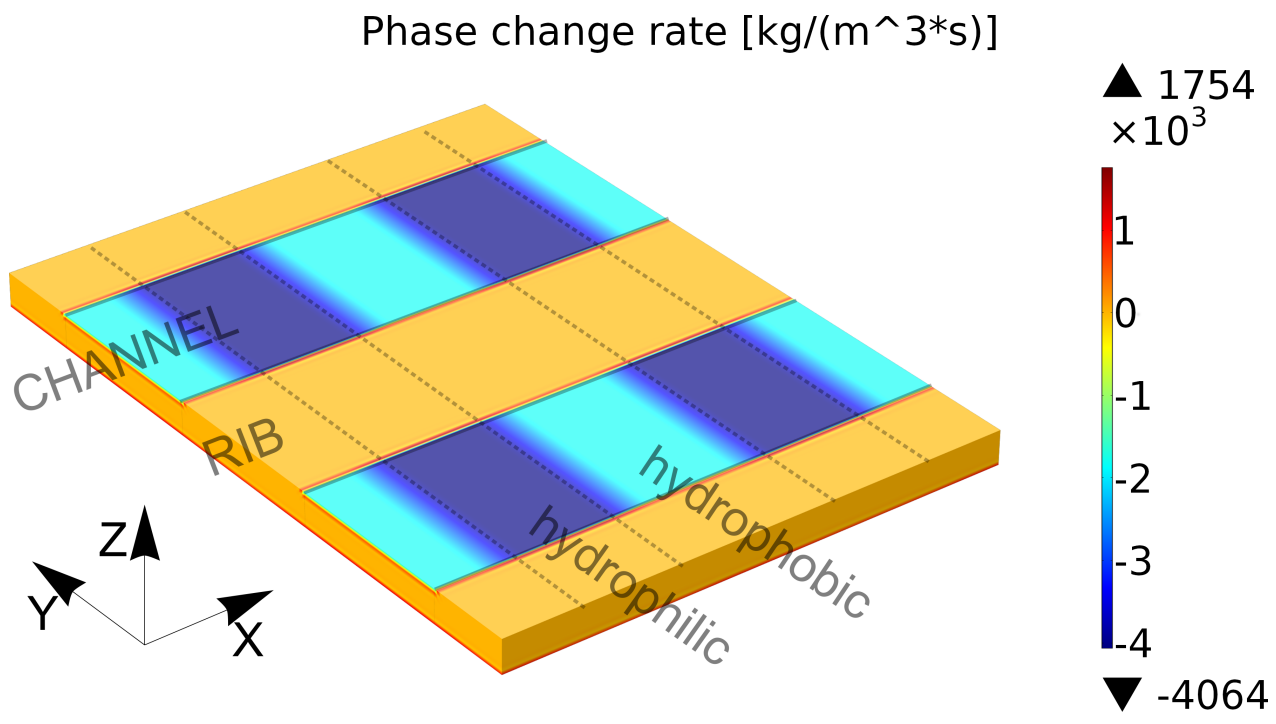

Figure 11: Simulated distribution of the phase change rates of the patterned GDL at a fuel cell voltage of 0.6 V. Positive values indicate the condensation process. Low humidity of the inlet gases causes the lowest rates (evaporation) under the channel regions. Different levels of saturation cause the rate difference between the hydrophobic and the hydrophilic regions under the channels. The condensation takes place under the ribs and in the vicinity of the CL boundary, which is related to higher water vapor concentration caused by the chemically produced water and the water dragged from the anode side.

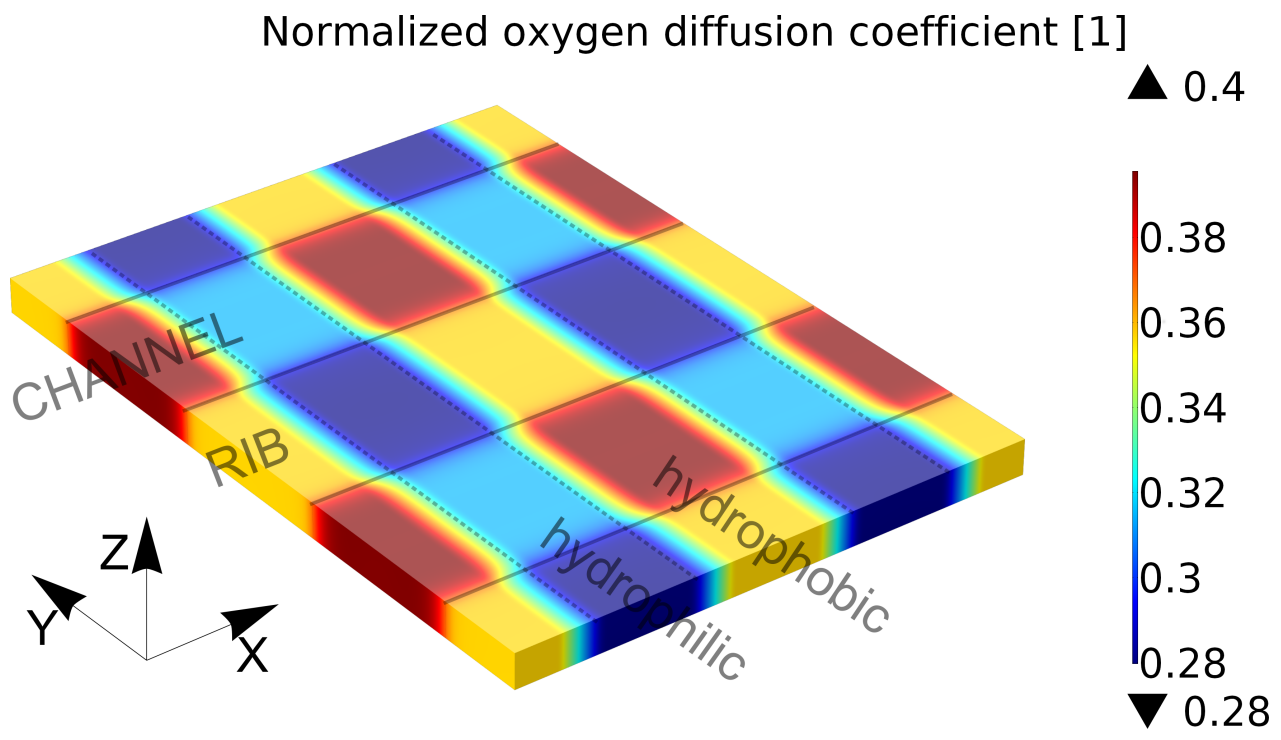

Figure 13: Simulated distribution of the normalized oxygen diffusion coefficient ($D_{O_2}/D_{O_2}^{int}$) of the patterned GDL at a fuel cell voltage of 0.6 V. Both the effective porosity and the liquid water saturation level influence the diffusion coefficient. The lowest value of 0.28 is observed under the ribs in the hydrophilic region, where the porosity is the lowest and the saturation is the highest. The highest value of 0.40 is found under the channels in the hydrophobic region, where the saturation is the lowest and the porosity is the highest.

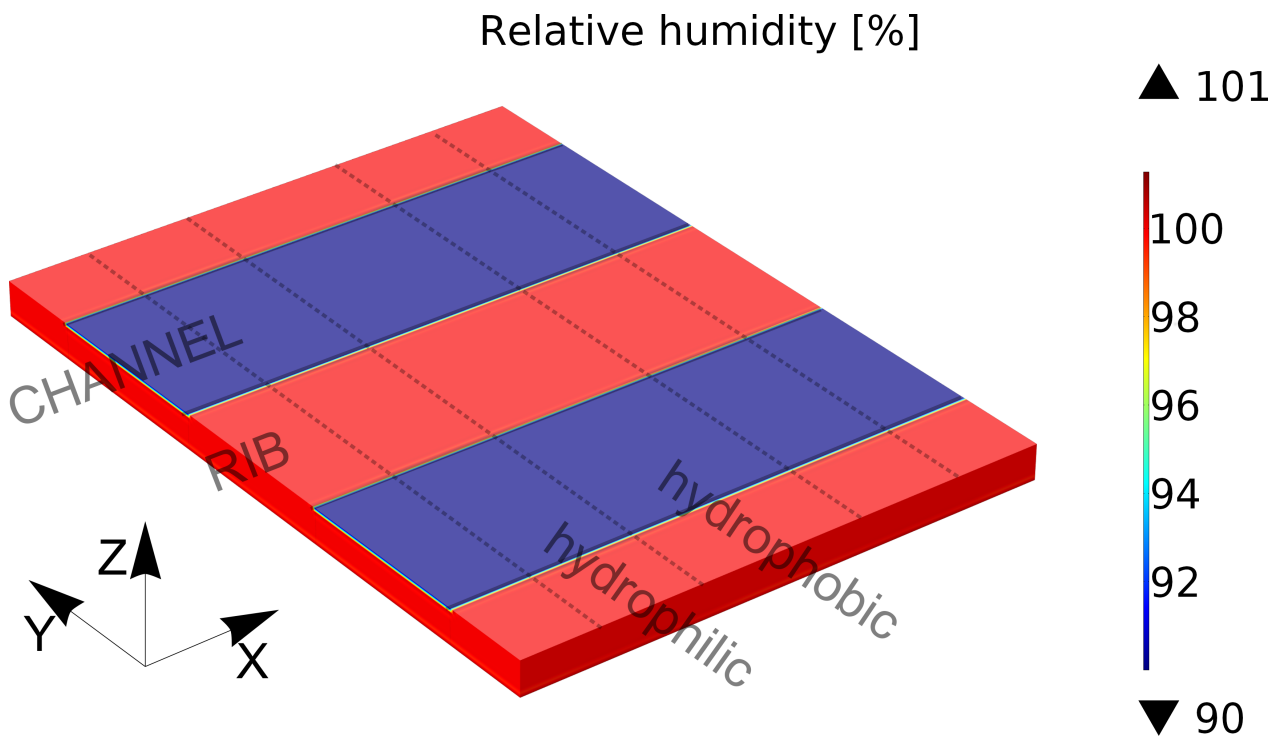

Figure 12: Simulated relative humidity of the patterned GDL at a fuel cell voltage of 0.6 V. The lowest values under the channels (90%) are related to the prescribed water vapor boundary condition. The highest values (above 100%) are observed in the vicinity of the CL boundary due to the electrochemical production and the transport of water vapor in the through-plane direction of the membrane.

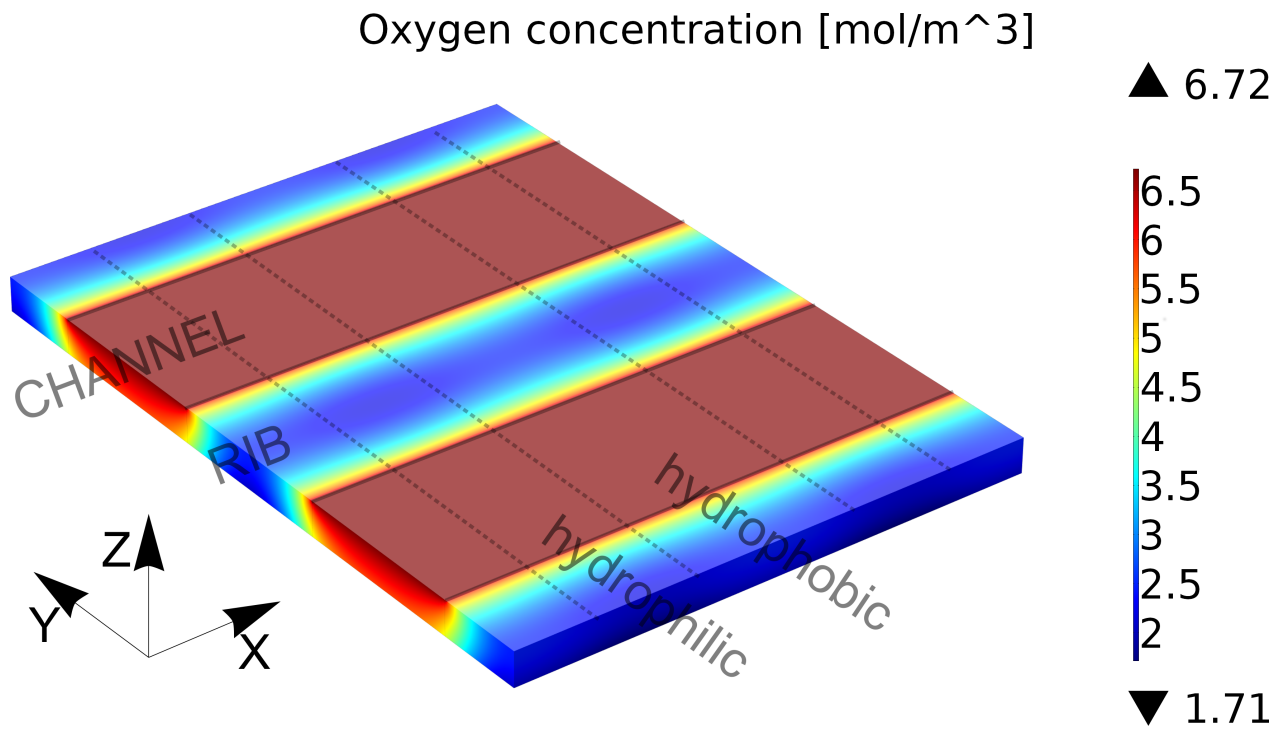

Figure 14: Top side view of the simulated oxygen concentration of the patterned GDL at a fuel cell voltage of 0.6 V. The highest value is observed under the channels (6.72 mol/m$^3$), where the oxygen concentration boundary condition is set. The lowest value on the top (bipolar plate) side is observed under the ribs in the hydrophilic region (2.45 mol/m$^3$).

In Fig. 12 we present the relative humidity distribution. The values are the lowest (90%) under the channels where the boundary water vapor condition is set. The highest values are close to the CL boundary (above 100%) where water vapor is produced. Moreover, water transported via electroosmotic drag from the anode side also contributes to higher values of relative humidity at this position. Back diffusion is considered in the simulation (see Eq. 25), however its effect is small due to the small gradient in relative humidity between the anode and the cathode side.

*Gas species transport.* – The species transport is governed by the effective porosity as well as the saturation level. This can be clearly seen in Fig. 13, where the normalized oxygen diffusion coefficient distribution is plotted. Note that the same plot is also valid for the water vapor diffusion. Under the ribs in the hydrophilic region we have the lowest value of 0.28. Here, the porosity is the lowest and saturation is the highest. In contrast we have the highest oxygen diffusion coefficient value of 0.40 under the channels in the hydrophobic region. Here, the saturation is the lowest and the porosity is the highest.

In Figs. 14 and 15 the oxygen concentrations are plotted from the top side (bipolar plate side) and from the bottom side (catalyst layer side) respectively. The concentration of oxygen is the highest on the bipolar plate side under the channels (approx. 6.72 mol/m$^3$) and from there it is transported under the ribs and towards the CL boundary as shown with the oxygen streamlines in Fig. 16. On the catalyst boundary, where oxygen is consumed in our model, we observe the highest value of 5.79 mol/m$^3$ in the hydrophobic part of the channel region. The concentration in the hydrophilic part is slightly lower at 5.60 mol/m$^3$. The lowest values of oxygen concentration are present under the ribs. Here, the values in the hydrophilic portion are 1.72 mol/m$^3$ while 2.07 mol/m$^3$ is the value seen in the hydrophobic part.

The water vapor concentration is plotted in Fig. 17. The lowest values are observed under the channels (4.13 mol/m$^3$) where water vapor concentration boundary condition is set.

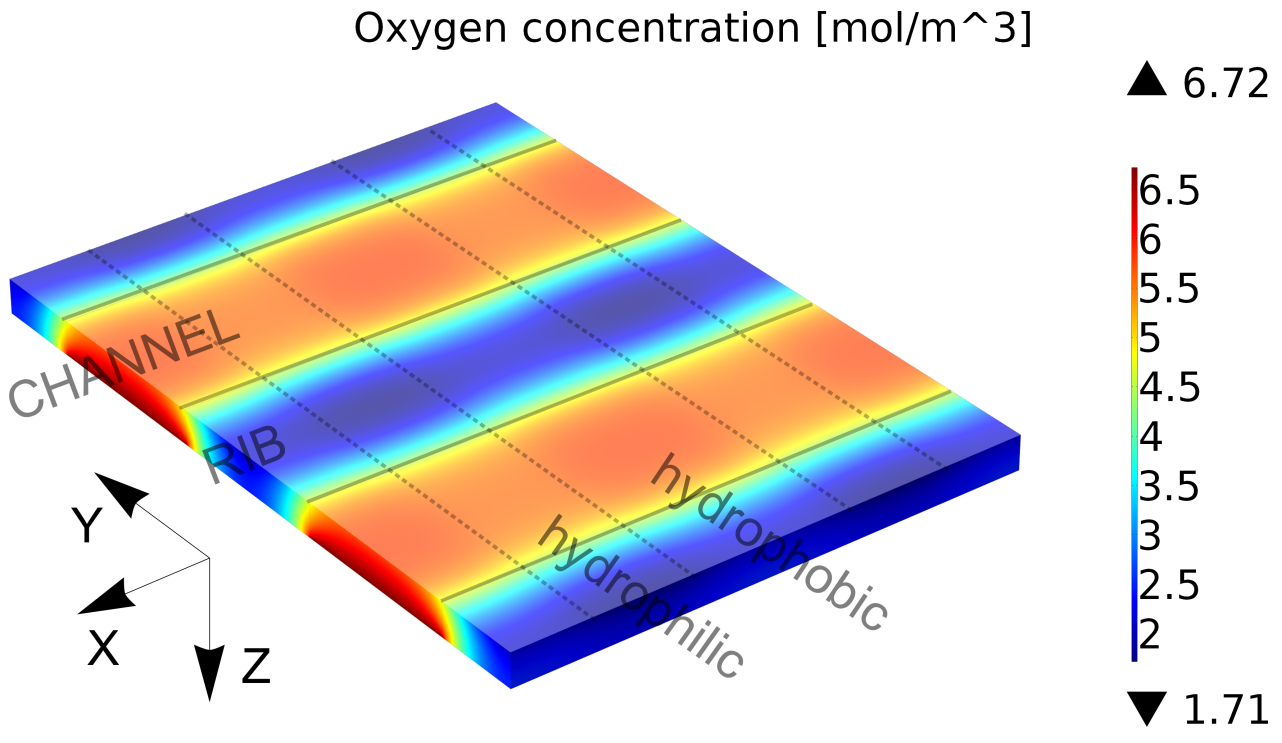

Figure 15: Bottom side view of the simulated oxygen concentration of the patterned GDL at a fuel cell voltage of 0.6 V. The highest value at the CL boundary is observed under the channels in the hydrophobic region ($5.79\,\mathrm{mol/m^3}$) and the lowest is under the ribs in the hydrophilic part ($1.72\,\mathrm{mol/m^3}$).

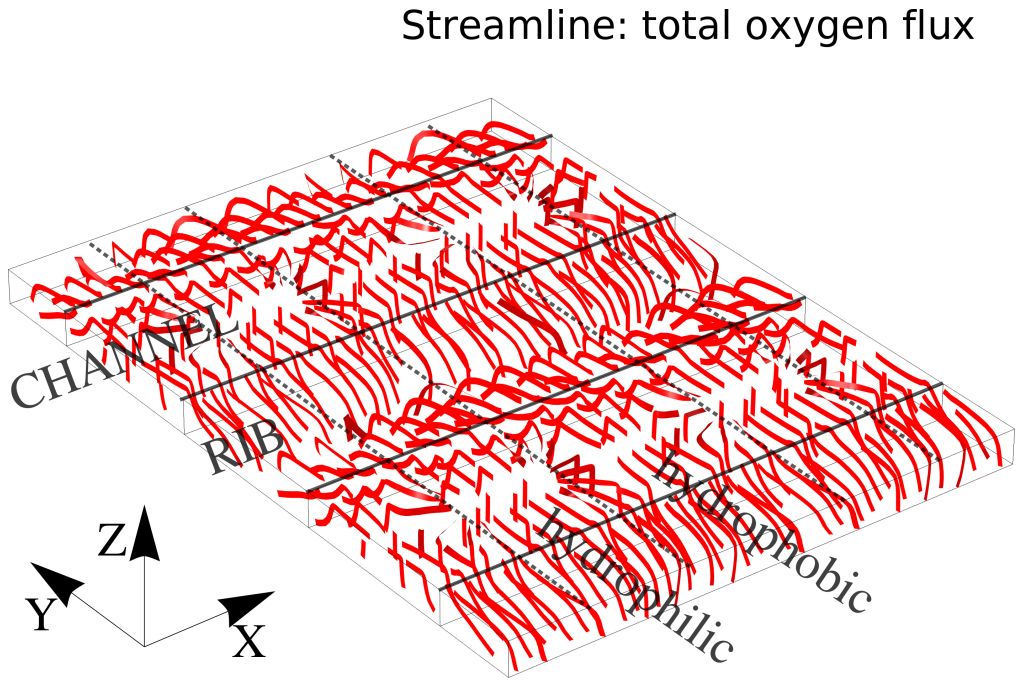

Figure 16: Simulated oxygen streamlines. The oxygen originates in the channel zones from which it is transported under the ribs and towards the CL boundary.

The highest values ($4.65\,\mathrm{mol/m^3}$) are observed in the vicinity of the CL boundary. This is caused by the electrochemical production and the through the membrane transport of the water vapor.

*Electrochemistry.* – Fig. 18 depicts the current density distribution at a fuel cell voltage of 0.6 V. We have the highest values below the channels where the current density varies from 1.68 $\mathrm{A/cm^2}$ in the hydrophilic region to 1.69 $\mathrm{A/cm^2}$ in the hydrophobic region. Lower values are observed under the ribs. Here the distribution varies from 1.35 $\mathrm{A/cm^2}$ in the hydrophilic region to 1.39 $\mathrm{A/cm^2}$ in the hydrophobic region.

In Fig. 19 we present the distribution of the total water vapor flux arising from the chemical reaction and the water transport through the membrane. In the current configuration the chemical reaction contributes one fourth and the electroosmotic drag approximately three quarters of the water flux. The contribution of back diffusion is due to very similar humidity conditions on the anode and the cathode side very small and in the range of $-2 \times 10^{-4}\,\mathrm{mol/m^3/s}$. The amount of chemically produced water and the amount of the through-membrane-dragged water are proportional to the current density (see Eq. 29). Hence, the water flux pattern is closely related to the current density pattern.

### 3.4 Parametric study

In this section we present the results of a parametric study. First, we show the influence of mesh size on the results.

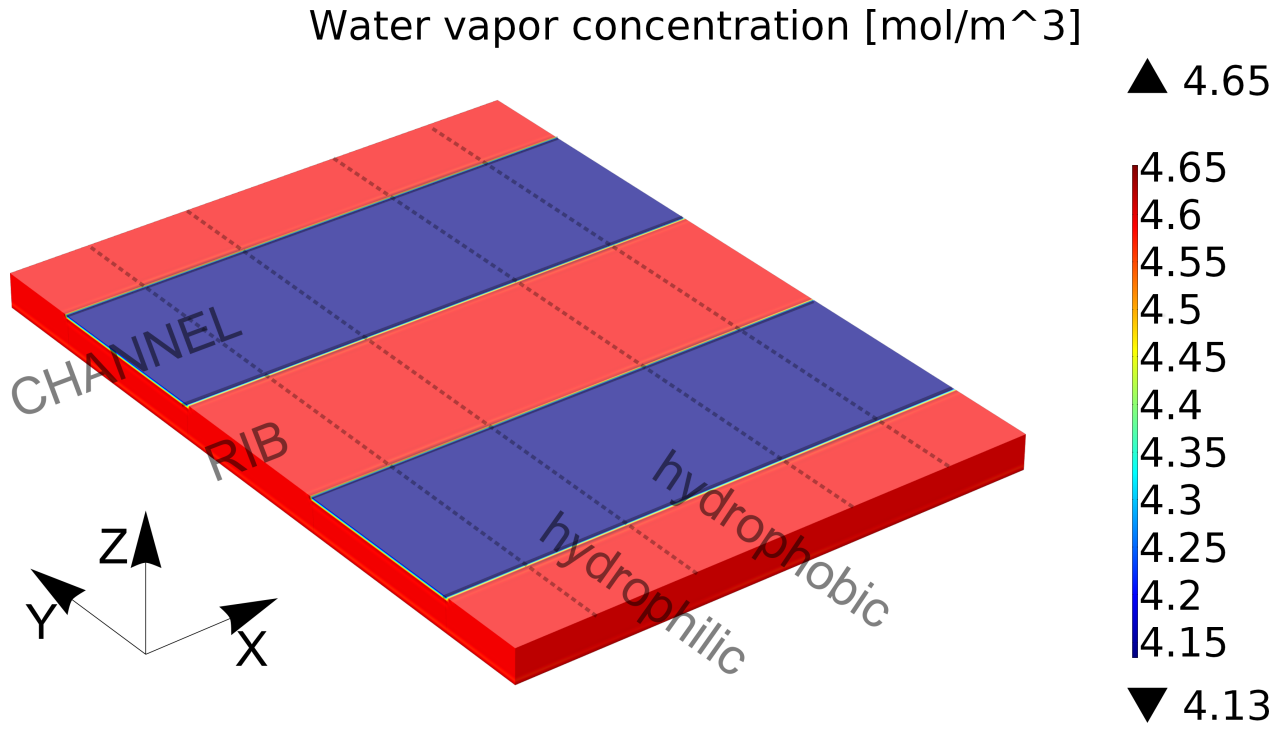

Figure 17: Simulated water vapor concentration of the patterned GDL at a fuel cell voltage of 0.6 V. The lowest values are observed under the channels (approx. $4.13\,\mathrm{mol/m^3}$) and are related to the prescribed water vapor concentration boundary condition. The highest values (approx. $4.65\,\mathrm{mol/m^3}$) are observed in the vicinity of the CL boundary due to the electrochemical production and the through the membrane transport of the water vapor.

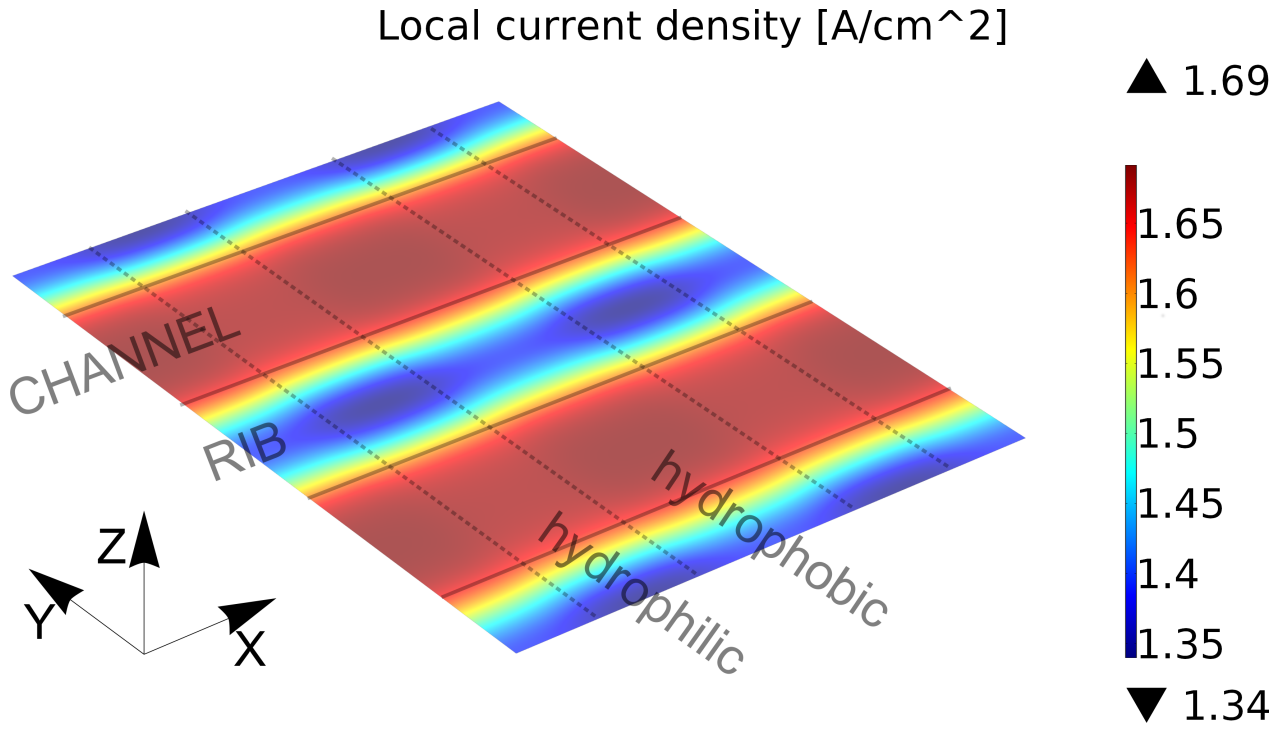

Figure 18: Simulated distribution of the current density of the patterned GDL at a fuel cell voltage of 0.6 V. The highest values are observed in the channel regions in the hydrophobic parts where the resistance to the transport of gases is the lowest. Contrarily, the lowest values are found in the ribs section in the hydrophilic region where the gas diffusion coefficient is the lowest.

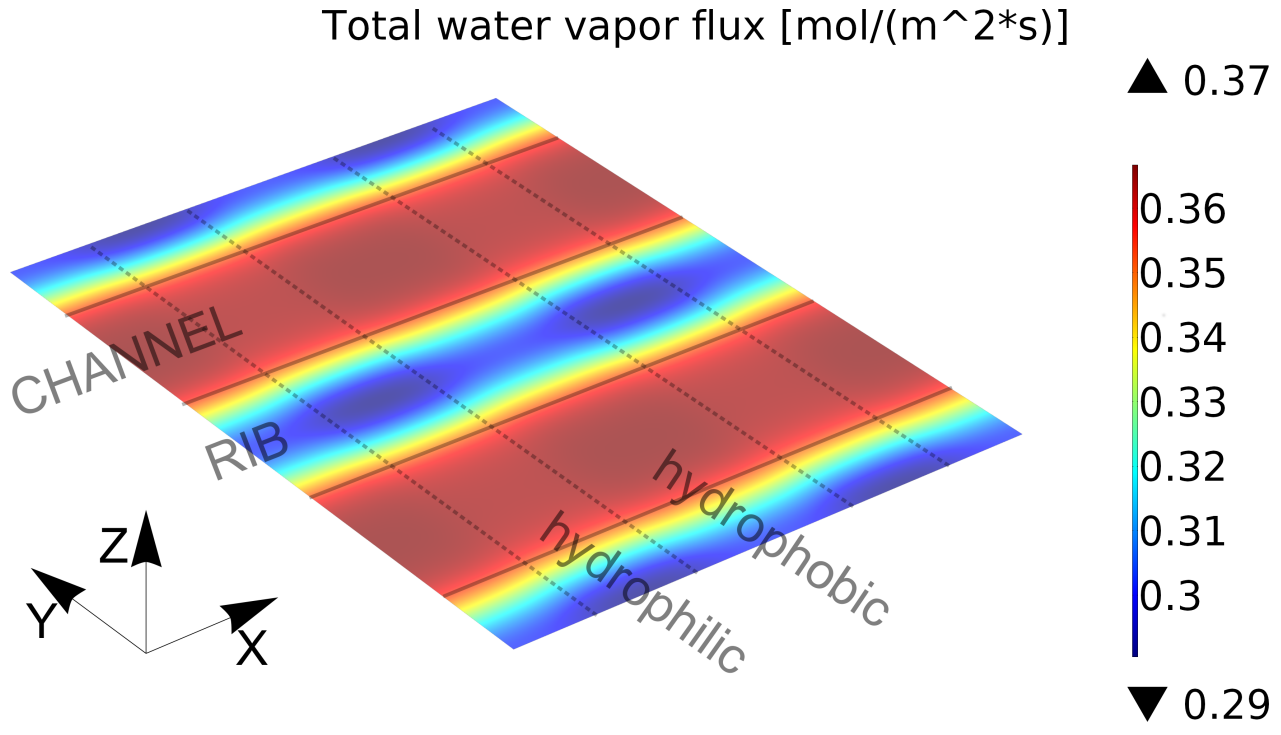

Figure 19: Simulated distribution of the water vapor flux of the patterned GDL at a fuel cell voltage of 0.6 V. The boundary flux at the CL/GDL interface includes the contribution of the chemical reaction in the catalyst layer as well as the contribution of the water transport through the membrane.

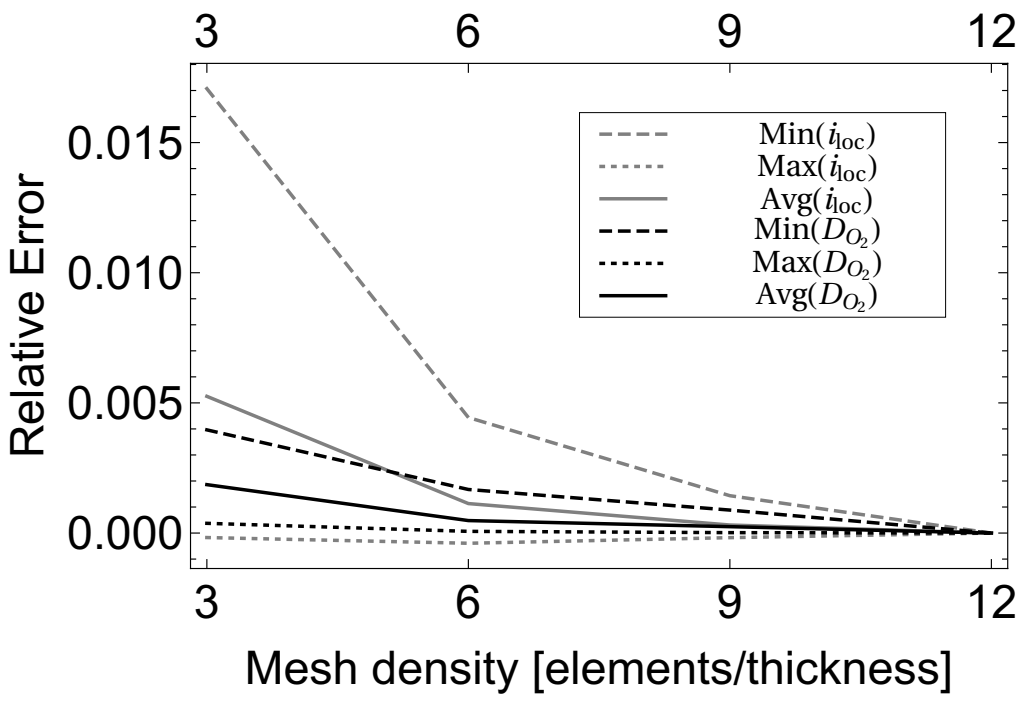

Figure 20: Relative errors of the local current density and the oxygen diffusion coefficient for different mesh sizes. The influence of mesh size has very little effect on the oxygen diffusion and the current density distribution. All values are below 1.6%.

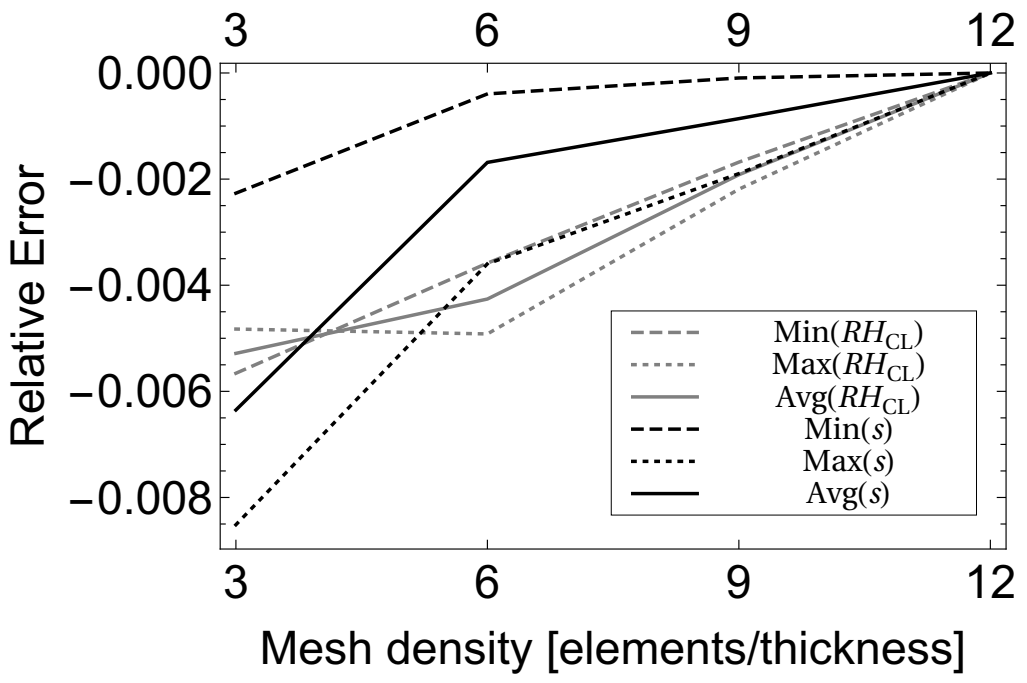

Figure 21: Relative errors of the relative humidity observed on the catalyst boundary and for the liquid water saturation. The results range between 0 and -0.8%. The largest error is observed for the maximal value of liquid water saturation when using the coarsest mesh.

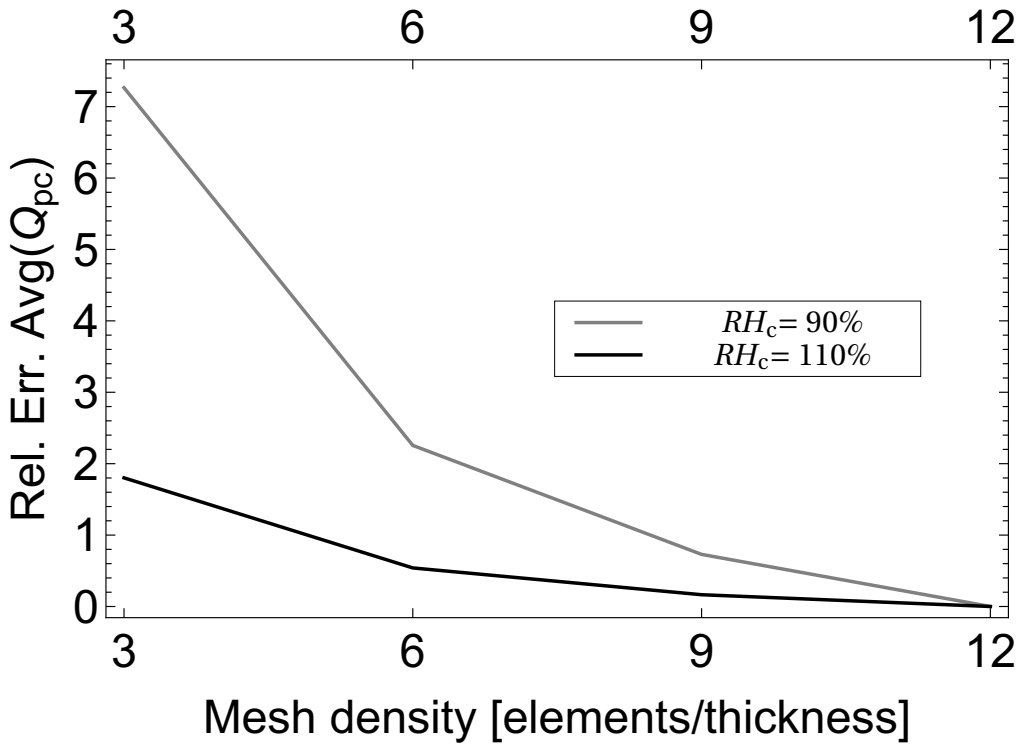

Figure 22: Relative errors of the average phase change mass transfer rate $Q_{\mathrm{pc}}$. The curves are obtained by considering the base case humidity ($\mathrm{RH_C} = 90\%$) and a case with higher humidity of cathode side ($\mathrm{RH_C} = 110\%$). Both curves show significant influence of mesh size on results. When both condensation and evaporation processes are present in one simulation (the base case scenario) the errors are larger.

Next, we investigate the sensitivity of our model to the relative humidity of cathode side inlet gases. Further, we perform a parametric sweep of interfacial area accommodation coefficient $\Gamma_{\mathrm{s}}$ which influences the phase change rates. Last, we show how the results vary if we use different values of prescribed capillary pressure which defines the Dirichlet boundary condition for saturation.

### 3.4.1 Sensitivity to mesh size

The influence of mesh size was examined by considering four different mesh densities. The coarsest meshed used has 3 elements in the thickness direction and a total of 288 cuboid elements. By doubling the number of elements in the thickness direction to 6 we obtained a mesh consisting of 3360 cuboids. Our second finest mesh, which is the one used to run simulations in Section 3.3 and which is shown in Fig. 7, has 9 elements in the thickness direction and a total of 12276 elements. The finest mesh has 12 elements in the through-plane direction and 29928 cuboid elements.

In Figs. 20, 21 and 22 we plot the relative error curves for minimal, maximal and average values of $i_{\mathrm{loc}}$, $D_{\mathrm{O_2}}$, $s$, the relative humidity observed on the catalyst layer boundary $RH_{\mathrm{CL}}$, and $Q_{\mathrm{pc}}$, where the finest mesh was considered as the benchmark which gives the most accurate results.

In Fig. 20 the error curves for the local current density and the oxygen diffusion coefficient are shown. We can see that the influence of mesh size is very small for the two parameters. The biggest observed error is approx. 1.6% when using the coarsest mesh and comparing the minimal values of local current density. The values for all other meshes lie below 0.5%.

Fig. 21 shows the curves of relative error for the relative humidity observed on the catalyst boundary and for the liquid water saturation. We can see that also here the results are not very sensitive to mesh size as all errors lie below 1%.

In Fig. 22 we plot the error curves for the phase change mass transfer rate $Q_{\mathrm{pc}}$. Here, the influence of mesh size is significant and the errors when using the coarsest mesh are 700% for the base case scenario ($\mathrm{RH_C} = 90\%$) and approx. 200% for the case with higher relative humidity of inlet gases ($\mathrm{RH_C} = 110\%$). Namely, the prefactor $\frac{A_{\mathrm{pore}}}{M_{\mathrm{w}}}\mathrm{Sh_c}\frac{D_{\mathrm{v}}}{d}$ of the evaporation/condensation rate (see Eq. (10)) is very large ($\approx 1.2 \times 10^8$ mol/kg/s) and even the smallest difference in the local value of relative humidity or the local level of liquid water saturation causes significant change in results. This effect is even more pronounced when using materials with different wettability properties. Here, we have to be sure that we use the size of the mesh that adequately describes the gradients in liquid water saturation.

### 3.4.2 Sensitivity to relative humidity

The influence of relative humidity of inlet gases was examined by varying the boundary value $\mathrm{RH_C}$ between 90 and 150%. In Figs. 23 and 24 the curves for several simulation results are plotted where the base case results, presented in Section 3.3, were used to normalize the values.

Fig. 23 shows that the influence on maximal, minimal and average value of oxygen diffusion coefficient is very small. All values of local current density drop with the increase of relative humidity. The effect is the largest on the minimal value of $i_{\mathrm{loc}}$ and it drops to approx. 95% of the base case value for $\mathrm{RH_C} = 150\%$.

Fig. 24 shows that the effect of relative humidity in the channels, at least in our testing range, has very little effect on the relative humidity at the catalyst layer and on the distributed values of liquid water saturation. The biggest observed effect is on the maximal and average values of $s$. They both increase by approx. 0.6% at $\mathrm{RH_C} = 150\%$.

### 3.4.3 Sensitivity to $\Gamma_{\mathrm{s}}$

The influence of the interfacial area accommodation coefficient $\Gamma_{\mathrm{s}}$ was tested by varying its value between 0.02 and 0.18. In Figs. 25 and 26 the curves for several simulation results are plotted where the base case results ($\Gamma_s = 0.1$) were used to normalize the values.

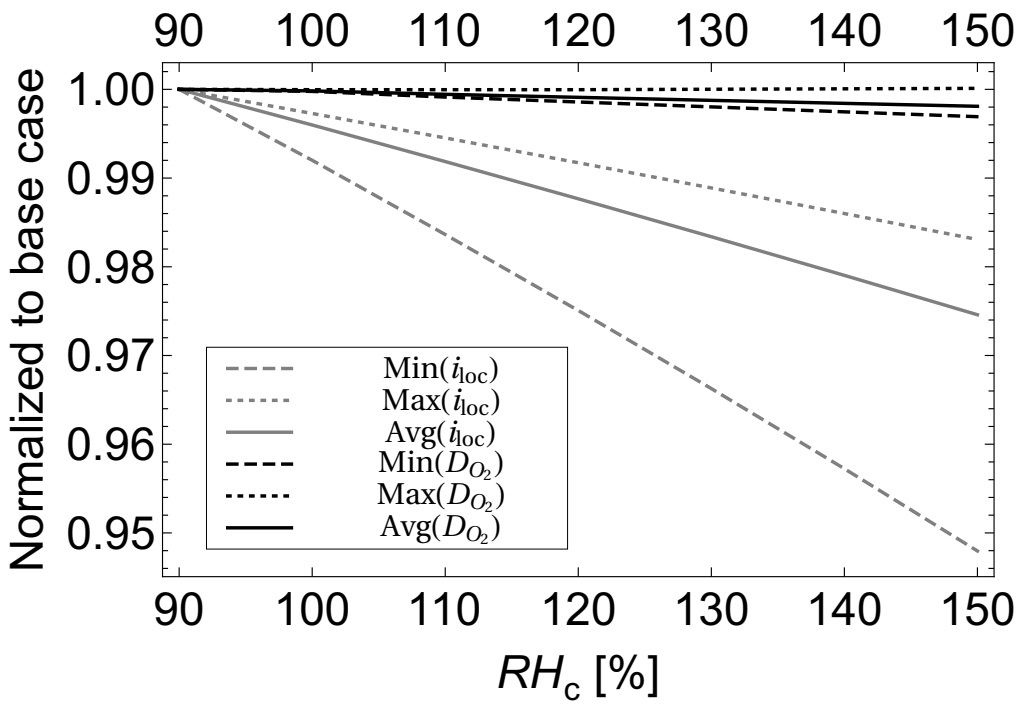

Figure 23: Local current density and oxygen diffusion coefficient versus relative humidity $RH_C$. The values are normalized to the base case ($RH_C = 90\%$) results.

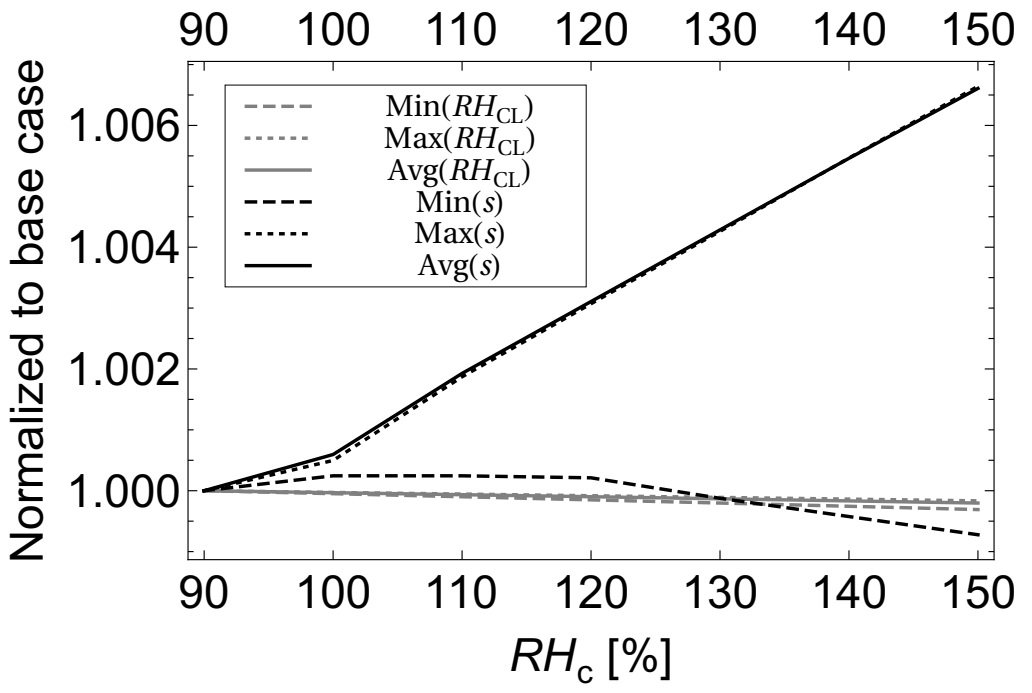

Figure 24: Relative humidity at the catalyst boundary and liquid water saturation versus relative humidity $RH_C$. The values are normalized to the base case ($RH_C = 90\%$) results.

We can see in Fig. 25 that $\Gamma_s$ has small direct influence on oxygen diffusivity and on local current density. In the high value range of $\Gamma_s$ the normalized values are between 0.9996 and 1.0004 while in the lower value range the spread increases between 0.9967 and 1.0024. The situation is similar when we take a look at the influence of $\Gamma_s$ on the relative humidity at the catalyst boundary and on the liquid water saturation in Fig. 26. At $\Gamma_s = 0.18$ all ratios lie between 0.9965 and 1.0001 and at $\Gamma_s = 0.02$ they range from 1.000 to 1.0254.

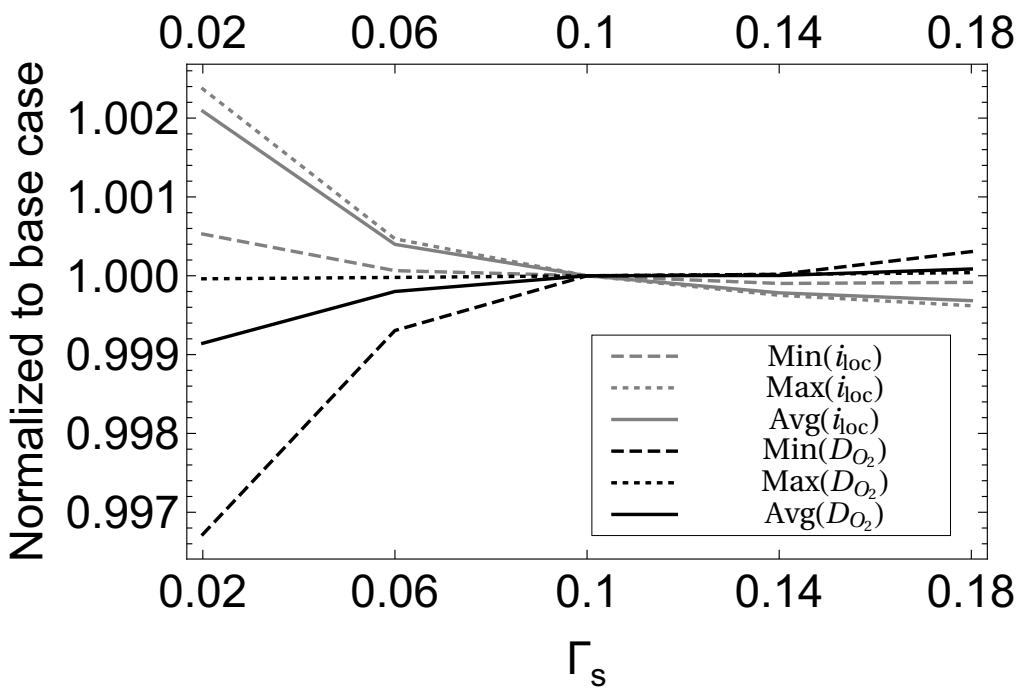

Figure 25: Local current density and oxygen diffusion coefficient as a function of the interfacial area accommodation coefficient $\Gamma_s$. The values are normalized to the base case results ($\Gamma_s = 0.1$).

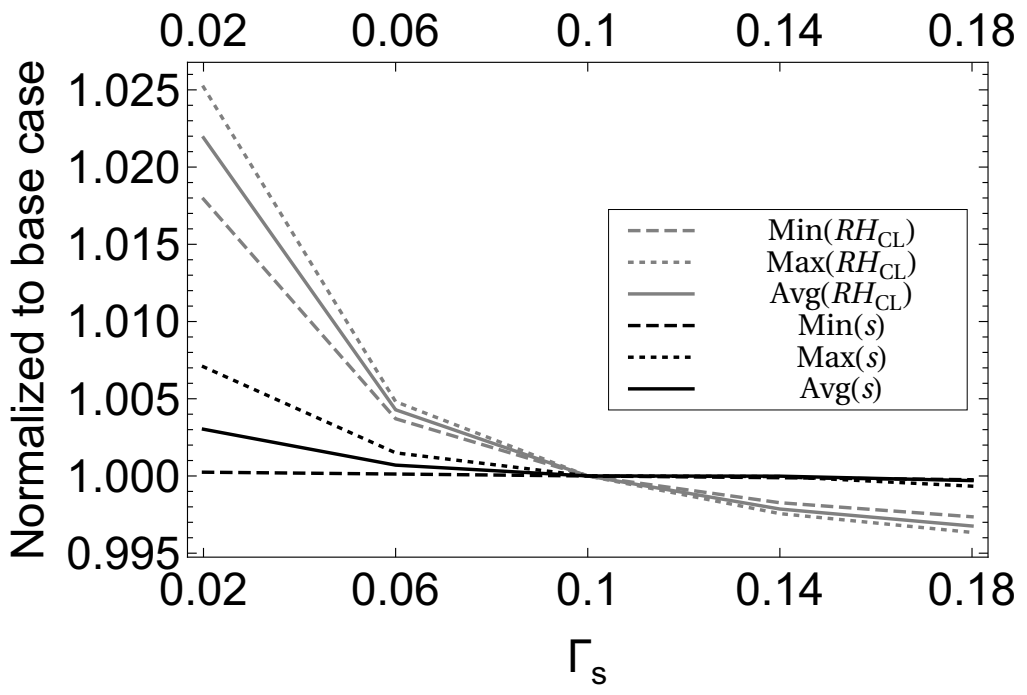

Figure 26: Relative humidity at the catalyst boundary and liquid water saturation as a function of the interfacial area accommodation coefficient $\Gamma_s$ curves. The values are normalized to the base case results ($\Gamma_s = 0.1$).

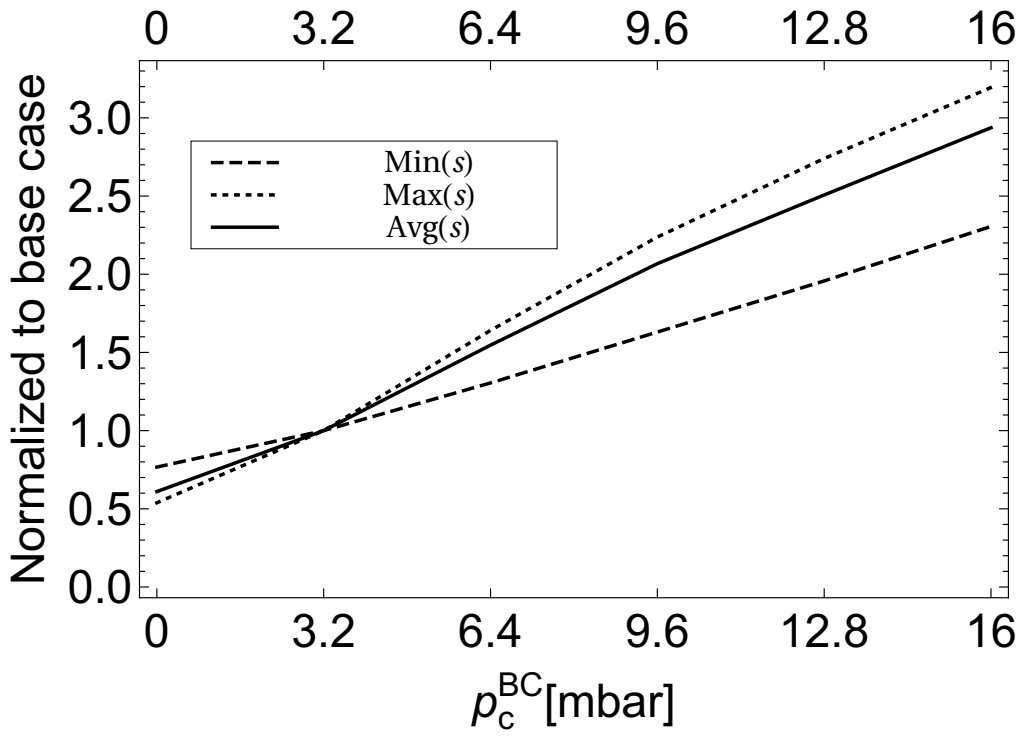

Figure 27: Liquid water saturation as a function of the boundary capillary pressure $p_c^{BC}$. The values are normalized to the base case results ($p_c^{BC} = 3.2$ mbar).

### 3.4.4 Sensitivity to applied boundary capillary pressure

Our last sensitivity study deals with the influence of applied boundary capillary pressure $p_c^{BC}$. By varying its value between 0 and 16 mbar we also directly influence the Dirichlet boundary condition for saturation at positions where the pressures of the two phases are defined. The results normalized to base case ($p_c^{BC} = 3.2$ mbar) are shown in Figs. 27 and 28.

As seen in Fig. 27, $p_c^{BC}$ has a significant effect on minimal, maximal and average values of $s$. Namely, the prescribed value of $p_c^{BC}$ defines the anchor point for $s$ while the evaporation/condensation processes define the variation from the anchor value.

From Fig. 28 it is evident that $p_c^{BC}$ has also a significant influence on the oxygen diffusion coefficient and through it on the local current density. Obviously, the higher the $p_c^{BC}$ the higher the saturation and the less available space for oxygen diffusion.

## 4 Conclusions and outlook

*Conclusions.* – A geometrically reduced steady-state numerical PEMFC model including the liquid water saturation of the GDLs was presented. It is based on a 3D representation of the cathode GDL, while the effects of the adjacent fuel cell components are treated as Dirichlet boundary conditions on the bipolar plate side and as boundary fluxes on the catalyst layer side, which depend on the solution of the local current density. The model includes a structural mechanics part simulating the influence of the mechanical assembly procedure, the Van Genuchten based two-phase mass transfer model including condensation/evaporation, and the electrochemical interface which also considers the transport of dissolved water between the anode and the cathode side.

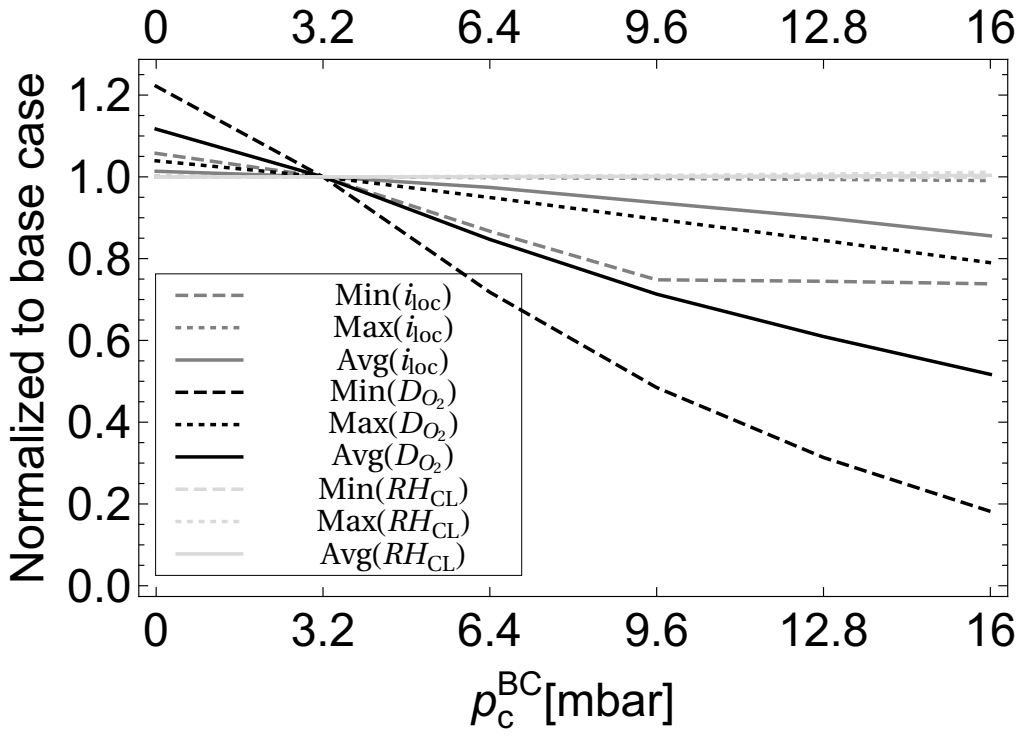

Figure 28: Local current density, oxygen diffusion coefficient, and relative humidity at the catalyst boundary versus the boundary capillary pressure $p_c^{BC}$. The values are normalized to the base case results ($p_c^{BC} = 3.2$ mbar).

The model has been applied to simulate an ex-situ imbibition process of a patterned GDL material. The simulation results have been compared to experimental data. It is concluded that the model can describe the GDL imbibition process with high accuracy. Both the experimental data as well as the model results show that the liquid water accumulates in the hydrophilic regions first and that higher capillary pressure is needed to saturate the hydrophobic regions.

Further, the model has been applied to simulate an operating cell. The simulated results are shown for a region consisting of two repetitions of hydrophilic/hydrophobic pattern and two repetitions of channel/rib pattern. The results show that both the channel/rib pattern and the hydrophilic/hydrophobic pattern influence the behavior of the fuel cell, confirming that the patterned design liberates the hydrophobic domains from liquid water and thus increases the local oxygen diffusivity.

Additionally, a parametric study for key simulation parameters has been performed. Most simulation results can be obtained with high accuracy even when using course meshes. However, when describing the condensation and evaporation processes, it is vital to use the mesh size that adequately describes the gradients in liquid water saturation. Surprisingly, the effect of $\Gamma_s$ is relatively small. We believe that this is caused by the very fast rate of condensation/evaporation process which, at least in our isothermal case, produces very similar relative humidity profiles for all values of $\Gamma_s$. As expected, the prescribed boundary value of saturation has a significant effect on all results of the operating fuel cell simulation.

*Outlook.* – The present model proved to be a valuable tool to study intricate couplings in an operating fuel cell and to examine the patterned GDL effects. However, there are several extensions to the model that we plan to pursue in order to develop the model to the predictive level where it could be used as an optimization tool. The setting of the appropriate saturation boundary conditions needs to be addressed. Determining more precise values of the phase change coefficients $\Gamma_m$ and $\Gamma_s$, will allow for more accurate description of the phase change rate distribution. Additionally, the inclusion of heat transport will provide the temperature distribution within the fuel cell.

## Acknowledgements

The authors gratefully acknowledge financial support from the Swiss Commission for Technology and Innovation in the framework of the Swiss Competence Center for Energy Research – Efficient Technologies and Systems for Mobility (SCCER Mobility). This work is part of the National Research Programme "Energy Turnaround" (NRP 70) of the Swiss National Science Foundation (SNSF Project No. 143432). Last but not least, the authors are also grateful to the Swiss Federal Office of Energy (BFE) for financial support.

## Nomenclature

| | |
|---|---|
| $c_{O_2}$ | oxygen concentration |
| $c_v$ | water vapor concentration |
| $\boldsymbol{C}$ | constitutive matrix |
| $D_{O_2}$ | effective oxygen diffusion coefficient |
| $D_{O_2}^{int}$ | intrinsic oxygen diffusion coefficient |
| $\overline{d}$ | characteristic length for water diffusion |
| $D_v$ | effective water vapor diffusion coefficient |
| $D_v^{int}$ | intrinsic water vapor diffusion coefficient |
| $D_\alpha$ | effective diffusion coefficient of species $\alpha$ |
| $D_\alpha^{int}$ | intrinsic diffusion coefficient of species $\alpha$ |
| $D_\lambda$ | diffusion coefficient of dissolved water |
| $F$ | Faraday's constant |
| $i_{loc}$ | local current density |
| $i_0$ | exchange current density |
| $j_{H_2}$ | molar flux of hydrogen |
| $j_{\alpha,conv}$ | convective flux species $\alpha$ |
| $j_{\alpha,diff}$ | diffusive flux species $\alpha$ |
| $j_\alpha$ | total flux species $\alpha$ |
| $j_\lambda$ | net flux of dissolved water |
| $K_{rel,liq}$ | relative permeability of liquid phase |
| $K_{rel,gas}$ | relative permeability of gas phase |
| $m$ | van Genuchten exponent |
| $M_w$ | Molar mass of water |
| $n$ | van Genuchten exponent |
| $n_{drag}$ | electroosmotic drag coefficient |
| $p_b$ | van Genuchten breakthrough pressure |
| $p_{liq}$ | pressure of liquid phase |
| $p_{O_2}$ | oxygen partial pressure |
| $p_{ref}$ | reference pressure |
| $p_{gas}$ | pressure of gas phase |
| $p_c$ | capillary pressure |
| $Q_{liq}$ | mass source/sink term of liquid phase |
| $Q_{O_2}$ | oxygen sink term |
| $Q_{pc}$ | phase change mass transfer rate |
| $Q_v$ | water vapor source/sink term |
| $Q_{gas}$ | mass source/sink term of gas phase |
| $Q_\alpha$ | source/sink term of species *alpha* |
| $R$ | universal gas constant |
| $s$ | liquid water saturation |
| $Sh_c$ | Sherwood number for condensation |
| $Sh_e$ | Sherwood number for evaporation |
| $T$ | temperature |
| $\boldsymbol{u}$ | displacement vector |
| $\boldsymbol{u}_{liq}$ | velocity vector of liquid phase |
| $\boldsymbol{u}_{gas}$ | velocity vector of gas phase |
| $V_m$ | molar volume of dry membrane |
| $\alpha$ | species index |
| $\delta_0$ | initial thickness of GDL |
| $\delta_c$ | compressed thickness of GDL |
| $\boldsymbol{\varepsilon}$ | strain tensor |
| $\epsilon_{eff}$ | effective porosity |
| $\eta_C$ | oxygen reduction reaction overpotential |
| $\eta_R$ | CCM resistance overpotential |
| $\epsilon_0$ | porosity of uncompressed GDL |
| $\epsilon_c$ | porosity of compressed GDL |
| $\epsilon_i$ | (Averaged) CCM ionomer volume fraction |
| $\varepsilon_v$ | volumetric strain |
| $\lambda$ | dissolved water content |
| $\lambda_A$ | value of $\lambda$ on anode side |
| $\lambda_C$ | value of $\lambda$ on cathode side |
| $\lambda^{eq}$ | equilibrated membrane water content |
| $\mu_{liq}$ | Dynamic viscosity of water |
| $\mu_{gas}$ | Dynamic viscosity of oxygen |
| $\boldsymbol{\sigma}$ | stress tensor |
| $\phi_{rev}$ | reversible cell potential |
| $\rho_{liq}$ | Mass density of liquid water |
| $\rho_{gas}$ | Mass density of gas mixture |
| $\overline{Q}_{O_2}$ | oxygen boundary flux |
| $\overline{Q}_v$ | water vapor boundary flux |
| $\overline{\lambda}$ | average membrane water content |
| $\rho_v$ | water vapor mass density |
| $\sigma_p$ | protonic conductivity of membrane |

## Parameters for the base case simulation

| | |
|---|---|
| $A_{pore} = 23.4\,m^2/cm^3$ | Specific pore surface area of GDL [43] |
| $d_{CL} = 5\,\mu m$ | Catalyst coating thickness |
| $d_m = 25.4\,\mu m$ | Membrane thickness [44] |
| $E = 6.3\,MPa$ | Young's modulus of GDL [45] |
| $i_0 = 2.5 \times 10^{-4}\,A/m^2$ | Cathode exchange current density |
| $K_{abs} = 37.4\,\mu m^2$ | Absolute permeability of GDL [46] |
| $k = 2$ | Van Genuchten exponent [23] |
| $M_m = 1.087\,kg/mol$ | Equivalent weight of membrane [44] |
| $m = 0.8$ | Van Genuchten exponent [23] |
| $p_{ref} = 1\,atm$ | Reference pressure [30] |
| $T_{ref} = 25°C$ | Reference temperature [30] |
| $\alpha_C = 0.5$ | Cathode symmetry factor (assumed) |
| $\Delta G_{ref} = -237.18\,kJ/mol$ | Reference Gibbs free energy [30] |
| $\Delta S_{ref} = -163.25\,J/mol\,K$ | Reference reaction entropy [30] |
| $\epsilon_0 = 0.78$ | Porosity of undeformed GDL [47] |
| $\epsilon_{i,CL} = 0.3$ | Catalyst coating ionomer volume fraction [48] |
| $\epsilon_{i,m} = 1$ | Membrane ionomer volume fraction |
| $\eta_{OC} = 30\,mV$ | Open circuit overpotential |
| $\Gamma_m = 0.006$ | Liquid-vapor mass transfer coefficient [24] |
| $\Gamma_s = 0.1$ | Interfacial area accommodation coefficient [24] |
| $\nu = 0.09$ | Poisson's ratio of GDL [45] |
| $\rho_{dry} = 1980\,kg/m^3$ | Density of the dry membrane [31] |
| $\tau = 1.3$ | Tortuosity of GDL [49] |